\documentclass[11pt,envcountsect,envcountsame]{llncs}
\usepackage{makeidx}
\usepackage{amsfonts}
\usepackage{amsmath}
\usepackage{amssymb}
\usepackage{amscd}
\usepackage{graphicx}
\usepackage{comment}
\usepackage{color}
\usepackage{array,amssymb,amsmath,amsbsy}
\usepackage{enumerate}
\usepackage[bottom]{footmisc}
\usepackage[all,cmtip]{xy}
\usepackage{hyperref}
\usepackage{wrapfig}
\usepackage{compress}
\usepackage{xspace}
\usepackage{a4wide}
\usepackage{wrapfig}
\usepackage{simplemargins}
\usepackage[colorinlistoftodos,backgroundcolor=blue!10,linecolor=red,bordercolor=red]{todonotes}
\setallmargins{1in}
\newenvironment{packed_enum}{
    \begin{enumerate}
        \setlength{\itemsep}{1pt}
        \setlength{\parskip}{0pt}
        \setlength{\parsep}{0pt}
}{\end{enumerate}}

\newenvironment{packed_itemize}{
    \begin{itemize}
        \setlength{\itemsep}{1pt}
        \setlength{\parskip}{0pt}
        \setlength{\parsep}{0pt}
}{\end{itemize}}

\makeatletter
\renewcommand*\l@author[2]{}
\renewcommand*\l@title[2]{}
\makeatletter
\newcommand{\nocontentsline}[3]{}
\newcommand{\tocless}[2]{\bgroup\let\addcontentsline=\nocontentsline#1{#2}\egroup}

\newcommand{\heading}[1]{\smallskip\par\noindent{\bf #1}}

\def\computationproblem#1#2#3#4{
	\begin{center}
	\fbox{\begin{tabular}{rp{#4}}
	{\bf Problem:\enspace}&#1\\
	{\bf Input:\enspace}&#2\\
	{\bf Output:\enspace}&#3\\
	\end{tabular}}
	\end{center}
}

\def\calA{{\mathcal A}} \def\calB{{\mathcal B}} \def\calC{{\mathcal C}}
  \def\calF{{\mathcal F}}

\def\calM{{\mathcal M}}  \def\calO{{\mathcal O}}
\def\calP{{\mathcal P}}  
  
\def\calV{{\mathcal V}}

  \def\frakL{{\mathfrak L}}

\def\cP{\hbox{\rm \sffamily P}\xspace}
\def\cNP{\hbox{\rm \sffamily NP}\xspace}
\def\cGI{\hbox{\rm \sffamily GI}\xspace}

\def\cXP{\hbox{\rm \sffamily XP}\xspace}
\def\cFPT{\hbox{\rm \sffamily FPT}\xspace}
\def\ccoAM{\hbox{\rm \sffamily coAM}\xspace}
\def\cLog{\hbox{\rm \sffamily LogSpace}\xspace}

\def\O{\mathcal{O}{}}

\def\int{\hbox{\rm \sffamily INT}\xspace}

\def\Aut{{\rm Aut}}

\def\sec{{\rm sec}}

\def\dih{\mathbb{D}}
\def\cyc{\mathbb{Z}}

\def\dotcup{\mathbin{\dot\cup}}

\def\Aut{{\rm Aut}}

\def\O{\calO}

\def\bo{\partial} 
\def\int{\mathring} 

\def\L{\frakL}

\def\gi{\textsc{GraphIso}\xspace}
\def\lgi{\textsc{ListIso}\xspace}
\def\laut{\textsc{ListAut}\xspace}

\def\bgi{{\sc \bfseries GraphIso}\xspace}

\spnewtheorem{result}{Result}{\bfseries}{\itshape}

\DeclareMathOperator{\tw}{tw}

\title{Graph Isomorphism Restricted by Lists\thanks{%
The first and second authors are supported by CE-ITI (P202/12/G061 of GA\v{C}R).}}

\author{Pavel Klav\'{\i}k\inst{1} \and Du\v{s}an Knop\inst{2} \and Peter Zeman\inst{2}}

\institute{Computer Science Institute,\\Faculty of Mathematics and Physics,\\Charles University in
Prague, Czech Republic.\\
    \texttt{klavik@iuuk.mff.cuni.cz}. \and
	Department of Applied Mathematics,\\Faculty of Mathematics and Physics,\\Charles University in
	Prague, Czech Republic.\\
    \texttt{\{knop,zeman\}@kam.mff.cuni.cz}.}

\begin{document}

\maketitle

\begin{abstract}
The complexity of \emph{graph isomorphism} (\gi) is a famous unresolved problem in theoretical
computer science. For graphs $G$ and $H$, it asks whether they are the same up to a relabeling of
vertices.  In 1981, Lubiw proved that \emph{list restricted graph isomorphism} (\lgi) is
\cNP-complete: for each $u \in V(G)$, we are given a list $\L(u) \subseteq V(H)$ of possible images
of $u$. After 35 years, we revive the study of this problem and consider which results for \gi
translate to \lgi.

\hskip 2em We prove the following: 1) When \gi is \cGI-complete for a class of graphs, it translates
into \cNP-completeness of \lgi. 2) Combinatorial algorithms for \gi translate into algorithms for
\lgi: for trees, planar graphs, interval graphs, circle graphs, permutation graphs, bounded genus
graphs, and bounded treewidth graphs.  3) Algorithms based on group theory do not translate: \lgi
remains \cNP-complete for cubic colored graphs with sizes of color classes bounded by 8.

\hskip 2em Also, \lgi allows to classify results for the graph isomorphism problem.  Some algorithms
are robust and translate to \lgi.  A fundamental problem is to construct a combinatorial
polynomial-time algorithm for cubic graph isomorphism, avoiding group theory. By the 3rd result,
\lgi is \cNP-hard for them, so no robust algorithm for cubic graph isomorphism exists, unless $\cP =
\cNP$.\\

\end{abstract}

For a dynamic structural diagram of our results, see the following website (supported Firefox and Google Chrome):
\url{http://pavel.klavik.cz/orgpad/list_isomorphism.html}
\bigskip

\tableofcontents

\newpage

\section{Introduction} \label{sec:introduction}

For graphs $G$ and $H$, a bijection $\pi : G \to H$ is called an \emph{isomorphism} if $uv \in E(G)
\iff \pi(u)\pi(v) \in E(H)$. The \emph{graph isomorphism problem} (\gi) asks whether there exists an
isomorphism from $G$ to $H$. It obviously belongs to \cNP, and no polynomial-time algorithm is
known. It is a prime candidate for an intermediate problem with complexity between \cP\ and
\cNP-complete.  There are threefold evidences that \gi is unlikely to be \cNP-complete: equivalence of
existence and counting~\cite{babai77,mathon1979note}, \gi belongs to \ccoAM, so the
polynomial-hierarchy collapses if \gi is \cNP-complete~\cite{goldreich1986,schoning1988graph}, and
\gi can be solved in quasipolynomial time~\cite{babai_quasipoly}.  For a survey,
see~\cite{babai1996automorphism}.

\subsection{Graph Isomorphism Problem for Restricted Graph Classes and Parameters}

\begin{figure}[b!]
\centering
\includegraphics[scale=1]{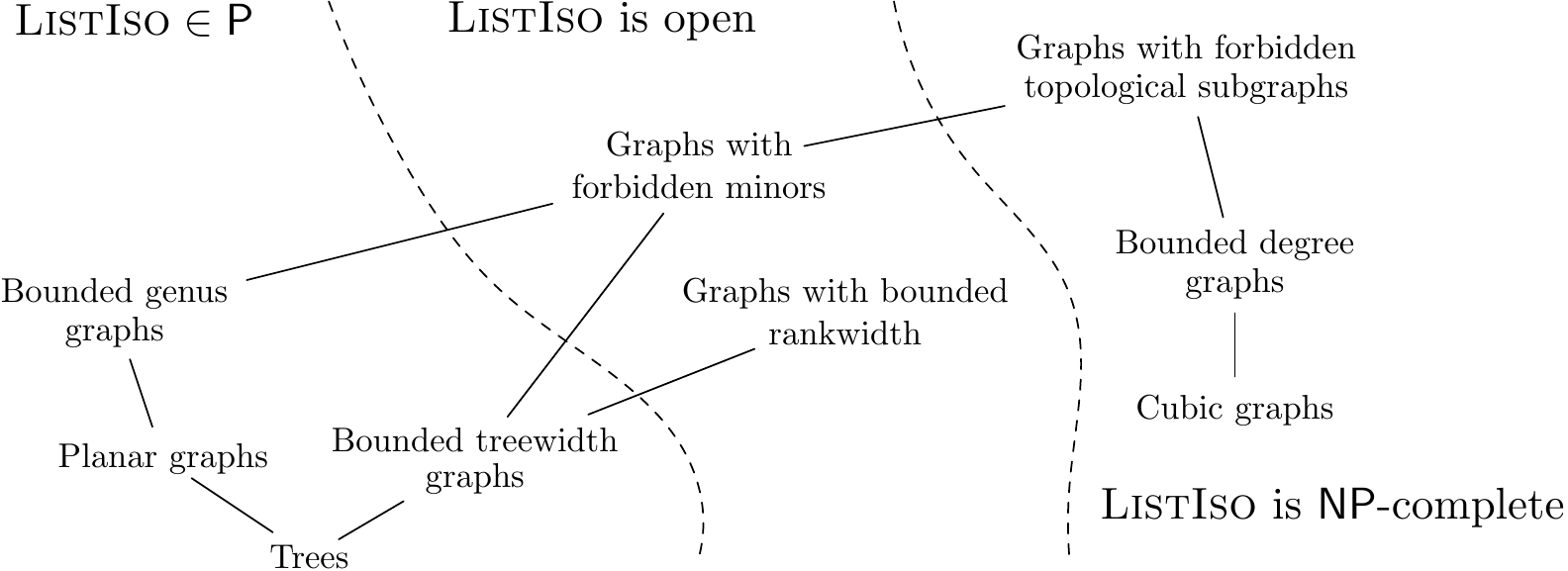}
\caption{Important graph classes for which the graph isomorphism problem can be solved in
polynomial time. Our complexity results for the list restricted graph isomorphism problem are
depicted.}
\label{fig:diagram}
\end{figure}

The graph isomorphism problem is solved efficiently for various restricted graph classes and
parameters, see Fig.~\ref{fig:diagram}.

\heading{Combinatorial Algorithms.}
A prime example is the linear-time algorithm for testing graph isomorphism of (rooted) trees. It is
a bottom-up procedure comparing subtrees. This algorithm is very robust and captures all possible
isomorphisms.  For many other graph classes, graph isomorphism reduces to graph isomorphism of
labeled trees: for planar
graphs~\cite{hopcroft_tarjan_dividing,hopcroft_tarjan_planar_iso,hopcroft1974linear}, interval
graphs~\cite{lueker1979linear}, circle graphs~\cite{hsu1995m}, and permutation
graphs~\cite{permutation_isomorphism,spinrad}.  Involved combinatorial arguments are used to solve
graph isomorphism for bounded genus graphs~\cite{pp_iso,filotti_mayer,miller_iso,kawarabayashi} and
bounded treewidth graphs~\cite{iso_xp_treewidth,iso_fpt_treewidth}.

\heading{Algorithms Based on Group Theory.}
The graph isomorphism problem is closely related to group theory, in particular to computing
generators of automorphism groups of graphs.  Assuming that $G$ and $H$ are connected, we can test
$G \cong H$ by computing generators of $\Aut(G \dotcup H)$ and checking whether there exists a
generator which swaps $G$ and $H$.  For the converse relation, Mathon~\cite{mathon1979note} proved
that generators of the automorphism group can be computed using $\O(n^3)$ instances of graph
isomorphism. 

Therefore, \gi can be attacked by techniques of group theory. A prime example is the seminal result
of Luks~\cite{luks1982isomorphism} which uses group theory to solve \gi for graphs of bounded degree
in polynomial time. If $G$ has bounded degree, its automorphism group $\Aut(G)$ may be arbitrary,
but the stabilizer $\Aut_e(G)$ of an edge $e$ is restricted. Luks' algorithm tests \gi by an
iterative process which determines $\Aut_e(G)$ in steps, by adding layers around $e$.

Group theory can be used to solve \gi of colored graphs with bounded sizes of color
classes~\cite{furst_hopcroft_luks} and of graphs with bounded eigenvalue
multiplicity~\cite{babai_eig_multiplicity,ponomarenko_eig_multiplicity}.
Miller~\cite{miller_k_contractible} solved \gi of $k$-contractible graphs (which generalize both
bounded degree and bounded genus graphs), and his results are used by
Ponomarenko~\cite{ponomarenko_minor} to show that \gi can be decided in polynomial time for
graphs with excluded minors. Luks' algorithm~\cite{luks1982isomorphism} for bounded degree graphs is also
used by Grohe and Marx~\cite{grohe_marx} as a subroutine to solve \gi on graphs with excluded
topological subgraphs.  The recent breakthrough of Babai~\cite{babai_quasipoly} heavily uses group
theory to solve the graph isomorphism problem in quasipolynomial time.

\heading{Is Group Theory Needed?} One of the fundamental problems for understanding the graph
isomorphism problem is to understand in which cases group theory is really needed, and in which
cases it can be avoided.\footnote{Ilya Ponomarenko in personal communication.} For instance, for
which graph classes can \gi be decided by the classical combinatorial algorithm called
$k$-dimensional Weisfieler-Leman refinement ($k$-WL)? (Described in Conclusions.)

Ponomarenko~\cite{ponomarenko_minor} used group theory to solve \gi in polynomial time on graphs
with excluded minors.  Robertson and Seymour~\cite{robertson_seymour} proved that a graph $G$ with
an excluded minor can be decomposed into pieces which are ``almost embeddable'' to a surface of genus
$g$, where $g$ depends on this minor. Recently, Grohe~\cite{grohe_graph_structure} generalized this
to show that for $G$, there exists a treelike decomposition into almost embeddable pieces which is
\emph{automorphism-invariant} (every automorphism of $G$ induces an automorphism of the treelike
decomposition). Using this decomposition, it is possible to solve graph isomorphism in polynomial
time and to avoid group theory techniques. In particular, $k$-WL can decide graph isomorphism on
graphs with excluded minors where $k$ depends on the minor.

It is a long-standing open problem whether the graph isomorphism problem for bounded degree graphs,
and in particular for cubic graphs, can be solved in polynomial time without group theory. For
instance, can cubic graph isomorphism be decided by $k$-WL for a suitable value $k$? Very recently,
fixed parameter tractable algorithms for graphs of bounded treewidth~\cite{iso_fpt_treewidth} and
for graphs of bounded genus~\cite{kawarabayashi} were constructed. On the other hand, the best known
parameterized algorithm for graphs of bounded degree is the \cXP algorithm of
Luks~\cite{luks1982isomorphism}, and it is a major open problem whether an \cFPT algorithm exists.

In this paper, we propose a different approach to show limitations of techniques used to attack the
graph isomorphism problem. We study its generalization called \emph{list restricted graph
isomorphism} (\lgi) which is \cNP-complete for general graphs. 

\begin{quote}
{\bf Implications for \bgi.} The study for \lgi allows to classify the results for the graph isomorphism problem.
An algorithm for \gi is called \emph{robust} if it
can be modified to solve \lgi while preserving the complexity. (Say, it remains a polynomial-time
algorithm, fixed parameter tractable algorithm, etc.)
\end{quote}

We show that combinatorial algorithms for graph isomorphism are robust. On the other hand, hardness
results for \lgi imply non-existence of robust algorithms for \gi. In particular, we show that \lgi is
\cNP-complete for cubic graphs, so no robust algorithm for cubic graph isomorphism exists, unless
$\cP = \cNP$. Similarly, no robust \cFPT algorithm for graph isomorphism of graphs of bounded degree
exists.

\subsection{List Restricted Graph Isomorphism}

\begin{figure}[t!]
\centering
\includegraphics[scale=1]{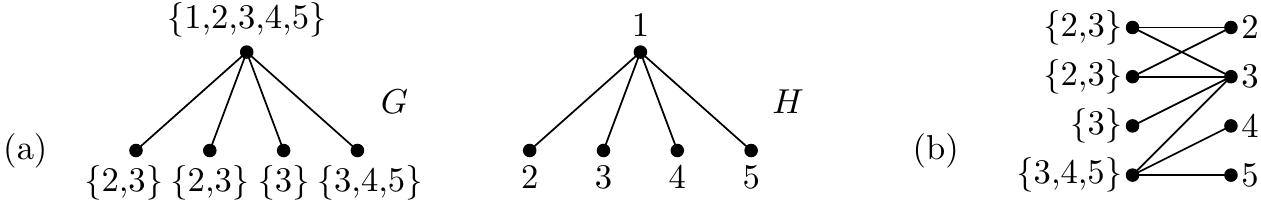}
\caption{(a) Two isomorphic graphs $G$ and $H$ with no list-compatible isomorphism.  (b) It does not
exist because there is no perfect matching between the lists of the leaves of $G$ and the leaves of
$H$.}
\label{fig:example}
\end{figure}

In 1981, Lubiw~\cite{lubiw1981some} introduced the following computational problems.  Let $G$ and
$H$ be graphs, and the vertices of $G$ be equipped with lists: each vertex $u \in V(G)$ has a
list $\frakL(u) \subseteq V(H)$. We say that an isomorphism $\pi : G \to H$ is
\emph{list-compatible} if, for all vertices $u \in V(G)$, we have $\pi(u) \in \frakL(u)$; see
Fig.~\ref{fig:example}a. A list-compatible isomorphism $\pi : G \to G$ is called a
\emph{list-compatible automorphism}. 

\computationproblem
{List restricted graph isomorphism -- \lgi}
{Graphs $G$ and $H$, and the vertices of $G$ are equipped by lists $\frakL(u) \subseteq V(H)$.}
{Is there a list-compatible isomorphism $\pi : G \to H$?}
{13cm}

\computationproblem
{List restricted graph automorphism -- \laut}
{A graph $G$ with vertices equipped with lists $\frakL(u) \subseteq V(G)$.}
{Is there a list-compatible automorphism $\pi : G \to G$?}
{10cm}

These two problems are polynomially equivalent (see Lemma~\ref{lem:lautlgi_equiv}).
Lubiw~\cite{lubiw1981some} proved the following surprising result:

\begin{theorem}[Lubiw \cite{lubiw1981some}] \label{thm:lubiw}
The problems \lgi and \laut are \cNP-complete.
\end{theorem}

\noindent Moreover, she proved that finding a fixed-point free involutory automorphism of a graph is
\cNP-complete.  Lalonde~\cite{lalonde} showed that it is \cNP-complete to decide whether a bipartite
graph has an involutory automorphism exchanging the parts; see~\cite{star_systems}.

Independently, \lgi was rediscovered in~\cite{fkkn,fkkn16b}. Given two graphs $G$ and $H$,
we say that $G$ \emph{regularly covers} $H$ if there exists a semiregular subgroup $\Gamma \le
\Aut(G)$ such that $G / \Gamma \cong H$. The list restricted graph isomorphism problem was used as a
subroutine in~\cite{fkkn,fkkn16b} for 3-connected planar and projective graphs to test regular
covering when $G$ is a planar graph. The key idea is that a planar graph $G$ can be reduced to a
3-connected planar graph $G_r$, for which $\Aut(G_r)$ is a spherical group.  Therefore, we can
compute all regular quotients $G_r / \Gamma_r$.  Next, we reduce $H$ towards $G_r / \Gamma_r$. The
problem is that subgraphs of $H$ may correspond to several different parts in $G$, so we compute
lists of all possibilities. One subroutine of the reduction leads to \lgi of 3-connected planar and
projective planar graphs, while the other leads to a generalization of bipartite perfect
matching~\cite{iv_matching}.

We note that other computational problems restricted by lists are frequently studied. \emph{List
coloring}, introduced by Vizing~\cite{list_coloring}, is \cNP-complete even for planar
graphs~\cite{list_coloring_planar} and interval graphs~\cite{list_coloring_int}. \emph{List
$H$-homomorphisms}, having a similar setting as \lgi, were also considered;
see~\cite{graphs_and_homomorphisms,list_hom_dichotomy,list_hom_removing}.

\subsection{Our Results} \label{sec:our_results}

We revive the study of list restricted graph isomorphism. The goal is to determine which techniques
for \gi translate to \lgi. We believe that \lgi is a very natural computational problem, as
evidenced by its application in~\cite{fkkn,fkkn16b}.  Further, its hardness results prove
non-existence of robust algorithms for the graph isomorphism problem itself. For instance, it is
believed that no \cNP-complete problem can be solved in quasipolynomial time. Therefore, some
techniques used by Babai~\cite{babai_quasipoly} to solve \gi in quasipolynomial time do not
translate to \lgi. To solve \gi efficiently, one necessarily has to apply such techniques.

The described algorithms for \lgi are a straightforward modification of previously known algorithms
for the graph isomorphism problem. The main point of this paper is \emph{not to develop new
algorithmic techniques}, but to \emph{classify known techniques for \gi from a different
viewpoint}. This viewpoint is the robustness of algorithms with respect to \lgi, and our results
give an insight into its meaning.

We prove the following three \emph{informal results} in this paper; see Fig.~\ref{fig:diagram} for
an overview:

\begin{result} \label{res:gi_completeness}
$ $\cGI-completeness results for \gi translate to \cNP-completeness for \lgi.
\end{result}

For many classes $\calC$ of graphs, it is known that \gi is equally hard for them as for general
graphs, i.e., it is \cGI-complete. For instance, \gi is \cGI-complete for bipartite graphs, split
and chordal graphs~\cite{lueker1979linear}, chordal bipartite and strongly chordal
graphs~\cite{strongly_chordal}, trapezoid graphs~\cite{trapezoid}, comparability graphs of dimension
4~\cite{kz}, grid intersection graphs~\cite{grid_GI_complete}, and line graphs~\cite{line_graphs}.
The polynomial-time reductions are often done in a way that all graphs are encoded into $\calC$, by
replacing each vertex with a small vertex-gadget.  (The constructions are quite simple, and the
non-trivial part is to prove that the constructed graph belongs to $\calC$.) As we prove in
Theorem~\ref{thm:vertex_gadget_reductions}, such reductions using vertex-gadgets also translate to
\lgi: they imply \cNP-completness of \lgi for $\calC$. For instance, \lgi is \cNP-complete for all
graph classes mentioned above (Corollary~\ref{cor:NPc_lgi}).

\begin{result} \label{res:combinatorial}
Combinatorial techniques for \gi translate to \lgi.
\end{result}

As a by-product, our paper gives a nice overview of the main combinatorial techniques involved in
attacking the graph isomorphism problem. These combinatorial techniques for \gi are often robust and
translate to \lgi straightforwardly.  Moreover, we can describe them more naturally with lists.

For example, the bottom-up linear-time algorithm for testing graph isomorphism of (rooted) trees
translates to \lgi in Theorem~\ref{thm:trees}, since it captures all possible isomorphisms. The key
difference is that the algorithm for \lgi finds perfect matchings in bipartite graphs, in order to
decide whether lists of several subtrees are simultaneously compatible; see Fig.~\ref{fig:example}b.
We use the algorithm of Hopcroft and Karp~\cite{hopcroft_karp}, running in time $\O(\sqrt n m)$.

The algorithms for graph isomorphism of planar, interval, permutation and circle graphs based on
tree decompositions and translate to \lgi, as we show in Theorems~\ref{thm:planar}, \ref{thm:int},
\ref{thm:perm}, and~\ref{thm:circle}. Even more involved algorithms for graphs isomorphism of
bounded genus and bounded treewidth graphs translate to \lgi in Theorems~\ref{thm:bounded_genus}
and~\ref{thm:fpt_treewidth}. The complexity for graphs with bounded rankwidth and graphs with
excluded minors remains open, see Conclusions for details.

\begin{result} \label{res:group_theory}
Group theory techniques for \gi do not translate to \lgi.
\end{result}

Group theory techniques do not translate to \lgi since list-compatible automorphisms of a graph $G$
do not form a subgroup of $\Aut(G)$. In Section~\ref{sec:group_theory}, when automorphism groups are
sufficiently rich, we show that \lgi remains \cNP-complete. In particular, we describe a non-trivial
modification of the original \cNP-hardness reduction of Lubiw~\cite{lubiw1981some} to show that \lgi
is \cNP-complete even for cubic colored graphs with color classes of size bounded by 16
(Theorem~\ref{thm:3reg_npc}). Therefore, no robust polynomial-time algorithm for cubic graph
isomorphism exists.

\section{Preliminaries and Outline}

Let $G$ be an input graph of \lgi or \laut.  We denote the set of its vertices by $V(G)$ and the set
of its edges by $E(G)$.  Let $n = |V(G)|$, $m = |E(G)|$ and $\ell$ be the total size of all lists.
To make the problem non-trivial, we can assume that $\ell \ge n$.

\heading{Bipartite Perfect Matchings.}
As a subroutine, we frequently solve \emph{bipartite perfect matching}:

\begin{lemma}[Hopcroft and Karp~\cite{hopcroft_karp}] \label{lem:bipmatch} The bipartite perfect
matching problem can be solved in time $\O(\sqrt{n}m)$, where $n$ is the number of vertices and $m$
is the number of edges.  \end{lemma}

\noindent For instance, when both $G$ and $H$ are independent sets, existence of a list-compatible
isomorphism is equivalent to existence of a perfect matching between the lists of $G$ and the
vertices of $H$. Therefore, using the algorithm of Hopcroft and Karp, we get the running time
$\O(\sqrt n \ell)$ while the input size is $\Omega(n+\ell)$. Finding bipartite perfect matchings is
the bottleneck in many of our algorithms and cannot be avoided: if it cannot be solved in linear
time, \lgi for many graph classes cannot be solved in linear time as well.

\heading{Outline: Main Points of This Paper.}
In Section~\ref{sec:basic_results}, we prove some basic results for \lgi such as polynomial-time
equivalence of \lgi and \laut and polynomial-time algorithms when maximum degree is 2 or all lists
are of size at most 2.

In Section~\ref{sec:gi_completeness}, we give a formal description of
Result~\ref{res:gi_completeness}.  We study polynomial-time reductions $\psi$ for \gi from a graph
class $\calC$ to another graph class $\calC'$: for each graph $G \in \calC$, the reduction $\psi$
produces another graph $G' \in \calC'$ such that $G \cong H$ if and only if $G' \cong H'$.  When
$\psi$ uses vertex-gadgets, it can be modified to a polynomial-time reduction for \lgi from $\calC$ to
$\calC'$. The vertex-gadget assumption means the following: in $G'$, each vertex $V(G)$ is replaced
by a small vertex-gadget while all automorphisms of $G'$ preserve vertex-gadgets and automorphisms
of $G$ induce automorphisms of $G'$ and vice versa.

In Section~\ref{sec:group_theory}, we give a formal description of Result~\ref{res:group_theory}. We
show that \lgi remains \cNP-complete for cubic colored graphs when each color class is of size at
most 16 and each list is of size at most 3. We modify the reduction of Lubiw~\cite{lubiw1981some} in
two steps. First, we reduce the problem from (positive) 1-in-3 SAT instead of 3-SAT. Therefore, only
positive literals appear and we reduce the sizes of lists from 7 to 3. Second, we modify variable
and clause gadgets to make the graph cubic.

In the remaining sections, we give a formal description of Result~\ref{res:combinatorial}.
In Section~\ref{sec:trees}, we modify the basic algorithm for graph isomorphism of (rooted) trees.
To deal with lists, we solve several bipartite perfect matching subroutines to test whether subtrees
are simultaneously list-compatible. The idea of this algorithm for \lgi of trees is used in some
other combinatorial algorithms.

In Section~\ref{sec:planar_graphs}, we describe that every planar graph can be decomposed into a tree
of its 3-connected components. Since \lgi can be easily solved on 3-connected planar graphs (using
geometry and uniqueness of embedding), we apply dynamic programming on this tree and solve \lgi for
general planar graphs as well.

In Section~\ref{sec:intersection}, we describe how to modify the algorithms for graph isomorphism of
interval, permutation and circle graphs. Similarly, they can be represented by MPQ-trees, modular
trees and split trees. Since we can solve \lgi on the graphs induced by nodes, we apply dynamic
programming on these trees and solve \lgi on interval, permutation and circle graphs as well.

In Section~\ref{sec:bounded_genus}, we modify the algorithm of Kawarabayashi~\cite{kawarabayashi}
to solve \lgi on graphs of bounded genus. This algorithm either uses a small number of possible
embeddings (which translates to \lgi), or finds a small cut of size at most 4 which is canonical,
splits the graph and test graph isomorphism of both pieces (which again translates to \lgi).

In Section~\ref{sec:bounded_treewidth}, we modify Bodlaender's \cXP
algorithm~\cite{iso_xp_treewidth} for graph isomorphism of graphs of bounded treewidth. The problem
is non-trivial since tree decomposition is not canonical. Therefore, it is a dynamic algorithm
running over all potential bags, and it can be easily modified with lists. Lokshtanov et
al.~\cite{iso_fpt_treewidth} obtain an \cFPT running time by computing a smaller set of potential
bags which is canonical.

In Section~\ref{sec:conclusions}, we conclude this paper with a group reformulation, related results
and open problems.

\section{Basic Results} \label{sec:basic_results}

In this section, we prove some basic results concerning the complexity of \lgi and \laut.

\begin{lemma} \label{lem:lautlgi_equiv}
Both problems \laut and \lgi are polynomially equivalent.
\end{lemma}

\begin{proof}
To see that \laut is polynomially reducible to \lgi just set $H$ to be a copy of $G$ and keep the
lists for all vertices of $G$. It is straightforward to check that these two instances are
equivalent.  For the other direction, we build an instance $G'$ and $\L'$ of \laut as follows. Let
$G'$ be a disjoint union of $G$ and $H$. And let $\L'(v) = \L(v)$ for all $v\in V(G)$ and set
$\L'(w) = V(G)$ for all $w \in V(H)$. It is easy to see that there exists  list-compatible
isomorphism from $G$ to $H$, if and only if there exists a list-compatible automorphism of $G'$.\qed
\end{proof}

\begin{lemma} \label{lem:list_size_two}
The problem \lgi can be solved in time $\O(n+m)$ when all lists are of size at
most two.
\end{lemma}

\begin{proof}
We construct a list-compatible isomorphism $\pi : G \to H$ by solving a 2-SAT formula which can be
done in linear time~\cite{even_2sat,aspvall_2sat}.  When $w \in \L(v)$, we assume that $\deg(v) =
\deg(w)$, otherwise we remove $w$ from $\L(v)$.  Notice that if $\L(u) = \{w\}$, we can set $\pi(u)
= w$ and for every $v \in N(u)$, we modify $\L(v) := L(v) \cap N(w)$. Now, for every vertex $u_i$
with $\L(u_i) = \{w^0_i, w^1_i\}$, we introduce a variable $x_i$ such that $\pi(u_i) = w^{x_i}_i$.
Clearly, the mapping $\pi$ is compatible with the lists.

We construct a 2-SAT formula such that there exists a list-compatible isomorphism if and only if it
is satisfiable. First, if $\L(u_i) \cap \L(u_j) \ne \emptyset$, we add implications for $x_i$ and
$x_j$ such that $\pi(u_i) \ne \pi(u_j)$. Next, when $\pi(u_i) = w^j_i$, we add implications that
every $u_j \in N(u_i)$ is mapped to $N(w^j_i)$. If $\L(u_j) \cap N(w^j_i) \ne \emptyset$, otherwise
$u_i$ cannot be mapped to $w^j_i$ and $x_i \ne j$. Therefore, $\pi$ obtained from a satisfiable
assignment maps $N[u]$ bijectively to $N[\pi(u)]$ and it is an isomorphism. The total number of
variables in $n$, and the total number of clauses is $\O(n+m)$, so the running time is
$\O(n+m)$.\qed
\end{proof}

\begin{lemma} \label{lem:disconnected}
Let $G_1,\dots,G_k$ be the components of $G$ and $H_1,\dots,H_k$ be the components of $H$.
If we can decide \lgi in polynomial time for all pairs $G_i$ and $H_j$, then we can
solve \lgi for $G$ and $H$ in polynomial time.
\end{lemma}

\begin{proof}
Let $G_1,\dots,G_k$ be the components of $G$ and $H_1,\dots,H_k$ be the components of $H$.  For each
component $G_i$, we find all components $H_j$ such that there exists a list-compatible isomorphism
from $G_i$ to $H_j$. Notice that a necessary condition is that every vertex in $G_i$ contains one
vertex of $H_j$ in its list. So we can go through all lists of $G_i$ and find all candidates $H_j$,
in total time $\O(\ell)$ for all components $G_1,\dots,G_k$.  Let $n' = |V(G_i)|$, $m' = |E(G_i)|$,
and $\ell'$ be the total size of lists of $G_i$ restricted to $H_j$. We test existence of a
list-compatible isomorphism in time $\varphi(n',m',\ell')$. Then we form the bipartite graph $B$
between $G_1,\dots,G_k$ and $H_1,\dots,H_k$ such that $G_iH_j \in E(B)$ if and only if there exists
a list-compatible isomorphism from $G_i$ to $H_j$.  There exists a list-compatible isomorphism from
$G$ to $H$, if and only if there exists a perfect matching in $B$.  Using Lemma~\ref{lem:bipmatch},
this can be tested in time $\O(\sqrt k \ell)$. The total running time depends on the running time of
testing \lgi of the components, and we note that the sum of the lengths of lists in these test is at
most $\ell$.\qed
\end{proof}

\begin{lemma} \label{lem:cycles}
The problem \lgi can be solved for cycles in time $\calO(\ell)$.
\end{lemma}

\begin{proof}
We may assume that $|V(G)| = |V(H)|$.  Let $u \in V(G)$ be a vertex with a smallest list and let $k
= |\L(u)|$.  Since $\ell = \calO(kn)$, it suffices to show that we can find a list-compatible
isomorphism in time $\calO(kn)$. We test all the $k$ possible mappings $\pi \colon G \to H$ with
$\pi(u) \in \L(u)$. For $u \in V(G)$ and $v \in \L(u)$, there are at most two possible isomorphisms
that map $u$ to $v$. For each of these isomorphism, we test whether they are list-compatible.\qed
\end{proof}

\begin{lemma} \label{lem:max_degree_2}
The problem \lgi can be solved for graphs of maximum degree 2 in time $\O(\sqrt n \ell)$.
\end{lemma}

\begin{proof}
Both graphs $G$ and $H$ are disjoint unions of paths and cycles of various lengths.  For each two
connected components, we can decide in time $\O(\ell')$  whether there exists a list-compatible
isomorphism between them, where $\ell'$ is the total size of lists when restricted to these
components: for paths trivially, and for cycles by Lemma~\ref{lem:cycles}. The rest follows from
Lemma~\ref{lem:disconnected}, where the running time is of each test in $\O(\ell')$ where $\ell'$ is
the total length of lists restricted to two components.\qed
\end{proof}

\section{GI-completeness Implies NP-completeness} \label{sec:gi_completeness}

Suppose that graph isomorphism is \cGI-complete for some class of graphs $\calC'$. We want to show
that in most cases, this translates in \cNP-completeness of \lgi for $\calC'$.

\heading{Vertex-gadget Reductions.}
Suppose that \gi is \cGI-complete for a class $\calC$. To show that \gi is \cGI-complete for another
class $\calC'$, one builds a polynomial-time reduction $\psi$ from \gi of $\calC$:
given graphs $G, H \in \calC$, we construct graphs $G',H' \in \calC'$ in
polynomial time such that $G \cong H$ if and only if $G' \cong H'$. We say that $\psi$ uses
\emph{vertex-gadgets}, if to every vertex $u \in V(G)$ (resp. $u \in V(H)$), it 
assigns a \emph{vertex-gadget} $\calV_u$, and these gadgets are 
subgraphs of $G'$ (resp. of $H'$), and satisfies the following two conditions:
\begin{packed_enum}
\item Every isomorphism $\pi : G \to H$ induces an isomorphism $\pi' : G' \to H'$ such that $\pi(u)
= v$ implies $\pi'(\calV_u) = \calV_v$.
\item Every isomorphism $\pi' : G' \to H'$ maps vertex-gadgets to vertex-gadgets and induces an
isomorphism $\pi : G \to H$ such that $\pi'(\calV_u) = \calV_v$ implies $\pi(u) = v$.
\end{packed_enum}

\begin{theorem} \label{thm:vertex_gadget_reductions}
Let $\calC$ and $\calC'$ be classes of graphs. Suppose that there exists a polynomial-time reduction
$\psi$ using vertex-gadgets from \gi of $\calC$ to \gi of $\calC'$.  Then there exists a
polynomial-time reduction from \lgi of $\calC$ to \lgi of $\calC'$.
\end{theorem}

\begin{proof}
Let $G,H \in \calC$ be an instance of \lgi. Using the reduction $\psi$, we construct the
corresponding graphs $G',H' \in \calC'$ with vertex-gadgets. We need to add lists for $V(G')$, we
initiate them empty. Let $u \in V(G)$. To all vertices $w$ of $\calV_u$, we add $\bigcup_{v \in
\L(u)} V(\calV_v)$ to $\L(w)$. For the vertices of $G'$ outside vertex-gadgets, we set the lists
equal to the union of all remaining vertices of $H'$.

We want to argue that there exists a list-compatible isomorphism $\pi' : G' \to H'$, if and only if
there exists a list-compatible isomorphism $\pi : G \to H$. If $\pi$ exists, by the first assumption
of the reduction, it induces $\pi'$ which is list-compatible by our construction of lists. On the
other hand, suppose that there exists a list-compatible isomorphism $\pi'$. By the second
assumption, $\pi'$ maps vertex-gadgets to vertex-gadgets and induces an isomorphism $\pi : G \to H$
which is list-compatible by our construction.\qed
\end{proof}

\begin{corollary} \label{cor:gi_completeness_translates}
Let $\calC$ be a class of graphs with \cNP-complete \lgi. Suppose that there exists a reduction
$\psi$ using vertex-gadgets from \gi of $\calC$ to \gi of $\calC'$. Then \lgi is \cNP-complete for
$\calC'$.\qed
\end{corollary}

Among others, this implies \cNP-completeness of \lgi for the following graph classes:

\begin{corollary} \label{cor:NPc_lgi}
The problem \lgi is \cNP-complete for bipartite graphs, split and chordal graphs, chordal bipartite
and strongly chordal graphs, trapezoid graphs, comparability graphs of dimension 4, grid intersection
graphs, and line graphs.
\end{corollary}

\begin{proof}
We use Corollary~\ref{cor:gi_completeness_translates} together with Theorem~\ref{thm:lubiw}.
We briefly describe \cGI-hardness reductions for every mentioned class. It is easy to check that,
except for line graphs, all these reductions use vertex-gadgets, where $\calV_u = \{u\}$ for every
$u \in V(G) \cup V(H)$.
\begin{packed_itemize}
\item \emph{Bipartite graphs.} Assuming the graphs are not cycles, we subdivide every edge in the
input graphs $G$ and $H$.
\item \emph{Split and chordal graphs~\cite{lueker1979linear}.} We subdivide every edge in $G$ and
$H$ and add the complete graphs on the original vertices.
\item \emph{Chordal bipartite and strongly chordal graphs~\cite{strongly_chordal}.} For bipartite
graphs $G$ and $H$, we subdivide all edges $e_i$ twice, by adding vertices $a_i$ and $b_i$, we add paths
of length three from $a_i$ to $b_i$, and we add the complete bipartite graph between $a_i$'s and
$b_i$'s.
\item \emph{Trapezoid graphs~\cite{trapezoid}.} For bipartite graphs $G$ and $H$, we subdivide every
edge and add the complete bipartite graph on the original vertices.
\item \emph{Comparability graphs of dimension at most 4~\cite{kz15}.} Assuming the graphs are not
cycles, we replace every edge in $G$ and $H$ by a path of length 8.
\item \emph{Grid intersection graphs~\cite{grid_GI_complete}.} For bipartite graphs $G$ and $H$, we
subdivide every edge twice and add the complete bipartite graph on the original vertices.
\item \emph{Line graphs~\cite{line_graphs}.} Assuming the graphs are not $K_3$ and $K_{1,3}$, we
consider $G'$ and $H'$ being the line graphs of $G$ and $H$. For every $u \in V(G)$, we put $\calV_u
= \{e : e \in E(G), u \in e\}$, and similarly for $u \in V(H)$. By Whitney Theorem~\cite{line_graphs}, $G
\cong H$ if and only if $G' \cong H'$, and it is easy to observe that it is a reduction using
vertex-gadgets.
\qed
\end{packed_itemize}
\end{proof}

\section{Group Theory Techniques Do Not Translate} \label{sec:group_theory}

Using group theory techniques, graph isomorphism can be solved in polynomial time for graphs of
bounded degree~\cite{luks1982isomorphism} and for colored graphs with color classes of bounded
size~\cite{furst_hopcroft_luks}. In this section, we modify the reduction of
Lubiw~\cite{lubiw1981some} to show that \lgi remains \cNP-complete even for 3-regular colored graphs
with color classes of size at most 16 and each list of size at most 3.

\subsection{Modifying the Reduction of Lubiw} \label{sec:lubiw}

We modify the original \cNP-hardness reduction of \lgi by Lubiw~\cite{lubiw1981some}. The original
reduction is from 3-SAT, but instead we use \emph{1-in-3 SAT} which is \cNP-complete by
Schaefer~\cite{1in3_sat}: all literals are positive, each clause is of size $3$ and a satisfying
assignment has exactly one true literal in each clause.  We show that an instance of 1-in-3 SAT can
be solved using \laut.

\begin{figure}[b!]
\centering
\includegraphics{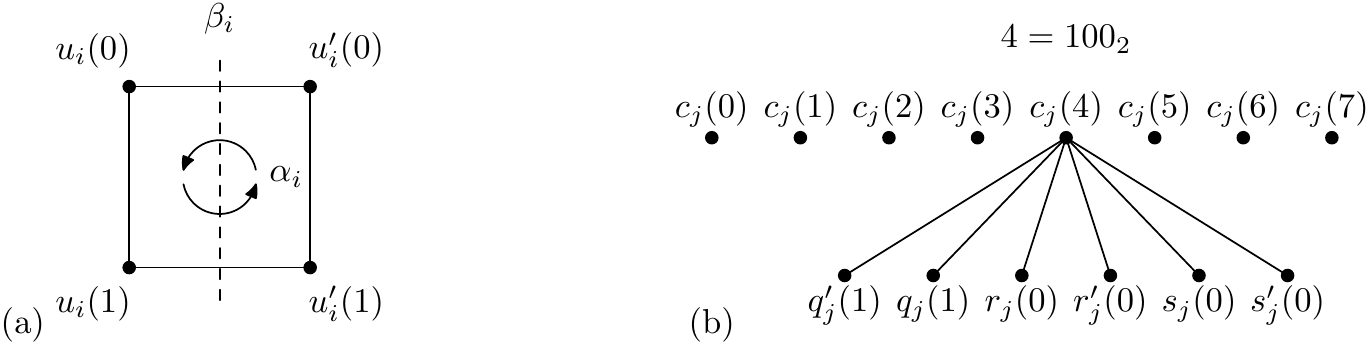}
\caption{Variable and clause gadgets.}
\label{fig:truth_gadget}
\end{figure}

\heading{Variable Gadget.}
For each variable $u_i$, we construct the \emph{variable gadget} $H_i$ which is a 4-cycle with the
vertices labeled as in Fig.~\ref{fig:truth_gadget}, and let $H$ be the disjoint union of these
cycles. Consider two automorphisms of $H_i$: the $180^{\circ}$ rotation $\alpha_i$ and the
reflection $\beta_i$. The automorphism $\alpha_i$ swaps $u_i(j)$ with $u'_i(1-j)$ while the
automorphism $\beta_i$ swaps $u_i(j)$ with $u'_i(j)$, for $j = 0,1$.  To a vertex $v \in V(H_i)$, we
assign the list $\L(v) = \{\alpha_i(v),\beta_i(v)\}$.

\heading{Clause Gadget.} Let $c_j$ be a clause with the literals $q_j$, $r_j$, and $s_j$.  For every
such clause $c_j$, the \emph{clause gadget} $G_j$ consists of the isolated vertices $c_j(0), \dots,
c_j(7)$. For every $k = 0,\dots,7$, we consider its binary representation $k = abc_2$, for $a,b,c \in
\{0,1\}$. We add an edge between $c_j(k)$ and the vertices $q_j(a), q_j'(a), r_j(b), r_j'(b),
s_j(c), s_j'(c)$ (these vertices belong to variable gadgets); see Fig.~\ref{fig:truth_gadget}b. We
assign the list $$\L(c_j(k)) = \{c_j(k \oplus 100_2), c_j(k \oplus 010_2), c_j(k \oplus 001_2)\},$$ where
$\oplus$ denotes the bitwise XOR; i.e., $\L(c_j(k))$ contains all $c_j(k')$ in which $k'$
differs from $k$ in exactly one bit.  Let $G$ be the resulting graph.

\begin{lemma} \label{lem:unique_extension}
Suppose that $\pi'$ is a partial automorphism obtained by choosing $\alpha_i$ or $\beta_i$ on each
variable gadget $H_i$. There exists a unique automorphism $\pi$ extending $\pi'$ such that
$\pi(G_j) = G_j$.
\end{lemma}

\begin{proof}
Let $c_j$ be a clause with the literals $q_j$, $r_j$, and $s_j$. We claim that $\pi(c_j(k))$ is
determined by the images of its neighbors.  Recall that $\beta_i$ preserves the numbers in brackets
of $H_i$, but $\alpha_i$ swaps them. Therefore, two neighbors of $\pi(c_j(k))$ are different from
the neighbors of $c_j(k)$ for every application of $\alpha_i$ on $q_j$, $r_j$ and $s_j$. Let $p =
abc_2$ such that $a = 1$, $b=1$ and $c=1$ if and only if $\alpha_i$ is applied on the variable
gadget of $q_j$, $r_j$, and $s_j$, respectively. Therefore, $\pi(c_j(k)) = c_j(k \oplus p)$;
otherwise $\pi$ would not be an automorphism.\qed
\end{proof}

\begin{lemma} \label{lem:reduction_correctness}
The 1-in-3 SAT formula is satisfiable if and only if there exists a list-compatible automorphism of $G$.
\end{lemma}

\begin{proof}
Let $T$ be a truth value assignment satisfying the input formula. We construct a list-compatible
automorphism $\pi$ of $G$. If $T(u_i) = 1$, we put $\pi|_{H_i} =
\alpha_i$, and if $T(u_i) = 0$, we put $\pi|_{H_i} = \beta_i$. By Lemma~\ref{lem:unique_extension},
this partial isomorphism has a unique extension to an automorphism $\pi$ of $G$. It is
list-compatible since $\pi(c_j(k)) = c_j(k \oplus p)$ and $p \in \{100_2,010_2,001_2\}$ (since $T$
satisfies the 1-in-3 condition). 

For the other implication, let $\pi$ be a list-compatible automorphism. Then $\pi|_{H_i}$ is either
equal $\alpha_i$, or $\beta_i$, which gives the values $T(u_i)$. By
Lemma~\ref{lem:unique_extension}, $\pi(c_j(k)) = c_j(k \oplus p)$ and since $\pi$ is a
list-compatible isomorphism, we have $p \in \{100_2,010_2,001_2\}$. Therefore, exactly one literal in
each clause is true, so all clauses are satisfied in $T$.\qed
\end{proof}

The described reduction is clearly polynomial, so we have established a proof of Theorem~\ref{thm:lubiw}.


\subsection{NP-hardness Proof}

For colored graphs, we require that automorphisms preserve colors.  By a simple modification of the
above reduction, we get the following: 

\begin{theorem} \label{thm:3reg_npc}
The problem \lgi is \cNP-complete for 3-connected colored graphs for which each color class is of size
at most 16 and each list is of size at most 3.
\end{theorem}

\begin{proof}
We modify the graph $G$ to a 3-regular graph. For a clause gadget $G_j$ representing $c_j$, we add
three new vertices $c_{j,q}(k)$, $c_{j,r}(k)$ and $c_{j,s}(k)$, each adjacent to $c_j(k)$ and two
vertices of the variable gadget of the literal $q_j$, $r_j$, and $s_j$, respectively; see
Fig.~\ref{fig:treeReplacement}b. For the newly added vertices, we set lists of size 3 compatibly
with $\L(c_j(k))$.

\begin{figure}[t!]
\centering
\includegraphics{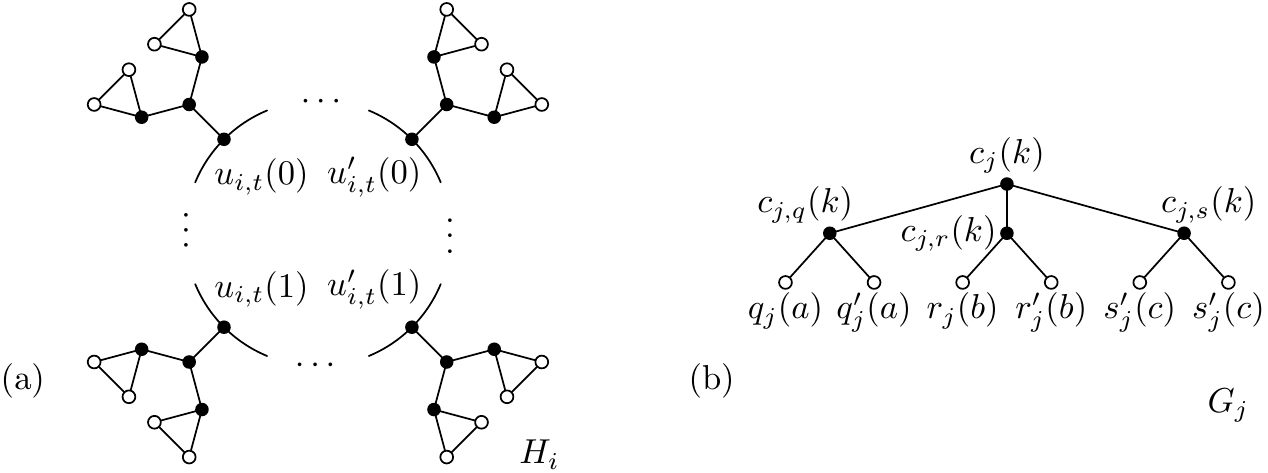}
\caption{The common vertices are depicted in white. (a) The variable gadget $H_i$. (b) The clause
gadget $G_j$.}
\label{fig:treeReplacement}
\end{figure}

\begin{figure}[b!]
\centering
\includegraphics{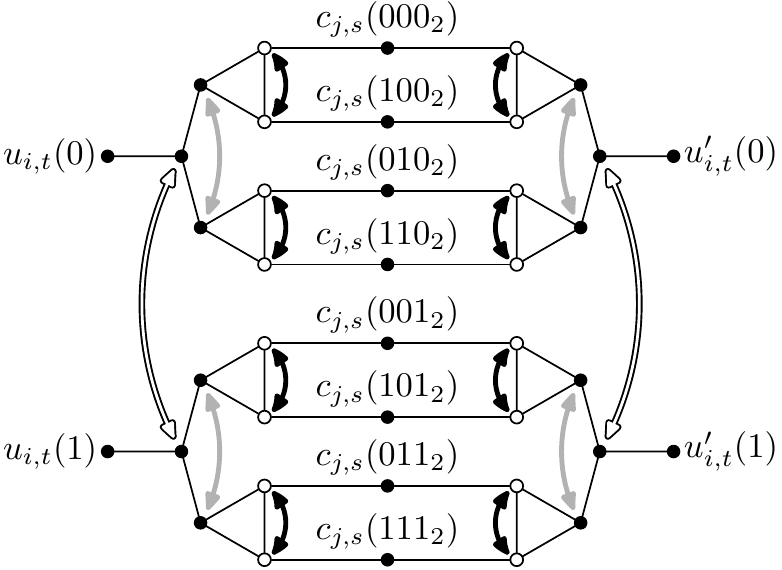}
\caption{Suppose that the variable $u_i$ is the literal $s_j$ of the clause $c_j$, so $k = yzx_2$.
We connect $H_i$ with $G_j$ as depicted. Suppose that an automorphism $\pi$
maps $c_{j,s}(k)$ to $c_{j,s}(k \oplus p)$. We depict the action of $\pi$ on the vertices of $H_i$
when $p = 001$ (in white), $p = 010$ (in gray), and $p = 100$ (in black).}
\label{fig:connecting_gadgets}
\end{figure}

Suppose that a variable $u_i$ has $o$ literals in the formula. We replace $H_i$
by a $4o$-cycle, together with a small gadget attached to each vertex as
depicted in Fig.~\ref{fig:treeReplacement}a.  Suppose that the vertices
$u_{i,1}(j),\dots,u_{i,o}(j)$ correspond to $u_i(j)$ and similarly
$u'_{i,1}(j),\dots,u'_{i,o}(j)$ correspond to $u'_i(j)$. The vertices of
$4o$-cycle are ordered:
$$u_{i,1}(0),\dots,u_{i,o}(0),u_{i,o}(1),\dots,u_{i,1}(1),u'_{i,1}(1),\dots,u'_{i,o}(1),
u'_{i,o}(0),\dots,u'_{i,1}(0).$$ We similarly define the automorphisms
$\alpha_i$ and $\beta_i$ and lists for the vertices on the cycle.

Consider the attached gadgets to four vertices corresponding to one literal of a clause $c_j$.  The
vertices depicted in white are adjacent to the added vertices $c_{j,u_i}(k)$ of $G_j$, as depicted
in Fig.~\ref{fig:connecting_gadgets}. Each $k$ consists of three bits, denoted $x$, $y$ and $z$ (in
some order). The bit $x$ corresponds to the literal (i.e, the first bit for $u_i = q_j$, and so on).
The white vertices of gadgets attached to $u_{i,t}(j)$ and $u'_{i,t}(j)$ are attached to
$c_{j,u_i}(k)$ where $x = j$.  Adjacent pairs of white vertices are connected to $c_j(k)$ where $k$
differs in another bit $y$.  Non-adjacent pairs of white vertices in one gadget are connected to
$c_j(k)$ where $k$ differs in the last bit $z$.

In Fig.~\ref{fig:connecting_gadgets}, the action of ${\mathbb Z}_2^3$ is depicted. When $\alpha_i$
or $\beta_i$ is used on $H_i$, they are further composed with the vertical reflection. Therefore,
Lemma~\ref{lem:unique_extension} translated to the modified definitions of variable and clause
gadgets which implies correctness of the reduction. The lists for the vertices of the attached
gadgets are created by composition of three depicted automorphisms together with $\alpha_i$ and
$\beta_i$; observe that they are of size at most $3$.

The constructed graph $G$ is 3-regular and all lists of $G$ are of size at most $3$.  We color the
vertices by the orbits of all list-compatible automorphisms and their compositions. Notice that each
color class is of size at most $16$. More precisely, all color classes on each $H_i$ are of size $16$,
and all color classes of clause gadgets $G_j$ are of size $8$.
\qed
\end{proof}

With Lemma~\ref{lem:max_degree_2}, we get a dichotomy for the maximum degree: \lgi can be
solved in time $\O(\sqrt n \ell)$ for the maximum degree 2, and it is \cNP-complete for the maximum
degree 3. Similarly, Lemma~\ref{lem:list_size_two} implies a dichotomy for the list sizes: \lgi can
be solved in time $\O(n+m)$ where all lists are of size 2, and it is \cNP-complete for lists of size
at most 3. For the last parameter, the maximum size of color classes, there is a gap.
Lemma~\ref{lem:list_size_two} implies that \lgi can be solved in time $\O(n+m)$ when all color
classes are of size 2 while it is \cNP-complete for size at most 16.

\section{Trees} \label{sec:trees}

In this section, we modify the standard algorithm for tree isomorphism to solve list restricted
isomorphism of trees. We may assume that both trees $G$ and $H$ are rooted, otherwise we root them
by their centers (and possibly subdivide the central edges). The algorithm for \gi process both
trees from bottom to the top. Using dynamic programming, it computes for every vertex possible
images using possible images of its children. This algorithm can be modified to \lgi.

\begin{theorem} \label{thm:trees}
The problem \lgi can be solved for trees in time $\O(\sqrt{n}\ell)$.
\end{theorem}

\begin{proof}
We apply the same approach with lists and update these lists as we go from bottom to the top.
After processing a vertex $u$, we compute an updated list $\L'(u)$ which contains all elements
of $\L(u)$ to which $u$ can be mapped compatibly with its descendants. To initiate, each leaf
$u$ of $G$ has $\L'(u) = \{w : \text{$w$ is a leaf and $w \in \L(u)$}\}$.

Next, we want to compute $\L'(u)$ and we know $\L'(u_i)$ of all children $U =
\{u_1,\dots,u_k\}$ of $u$.  For each $w \in \L(u)$ with $k$ children $w_1,\dots,w_k$, we want to
decide whether to put $w \in \L'(u)$.  Let $W = \{w_1,\dots,w_k\}$. Each $u_i$ can be mapped to
all vertices in $\L'(u_i) \cap W$. We need to decide whether all $u_i$'s can be mapped
simultaneously.  Therefore, we form a bipartite graph $B(U,W)$ between $U$ and $W$: we put an edge
$u_iw_j$ if and only if $w_j \in \L'(u_i)$. Simultaneous mapping is possible if and only if there
exists a perfect matching in this bipartite graph.

Let $r$ be the root of $G$ and $r'$ be the root of $H$.  We claim that there is a list-compatible
isomorphism $\pi : G \to H$, if and only if $\L'(r) = \{r'\}$. Suppose that $\pi$ exists. When
$\pi(u) = w$, its children $U$ are mapped to $W$. Since this mapping is compatible with the lists,
$w \in \L(u)$, and the mapping of $u_1,\dots,u_k$ gives a perfect matching in $B(U,W)$.
Therefore, $w \in \L'(u)$, and by induction $r' \in \L'(r)$. On the other hand, we can construct
$\pi$ from the top to the bottom. We start by putting $\pi(r) = r'$. When $\pi(u) = w$, we map its
children $U$ to $W$ according to some perfect matching in $B(U,W)$ which exists from the fact that
$w \in \L'(u)$.

It remains to argue details of the complexity. We process the tree which takes time $\O(\ell)$
(assuming $n \le \ell$) and we process each list constantly many times which takes $\O(\ell)$.
Suppose that we want to compute $\L'(u)$. We consider all vertices $w^1,\dots,w^p \in \L(u)$, and
let $W^j$ be the children of $w^j$.  We go through all lists of $\L'(u_1),\dots,\L'(u_k)$ in linear
time, and split them into sublists $\L'(u_i^j)$ of vertices whose parent is $w^j$. Only these
sublists are used in the construction of the bipartite graph $B(U,W^j)$.  Using
Lemma~\ref{lem:bipmatch}, we decide existence of a perfect matching in time $\O(\sqrt{k}\ell_j)$
which is at most $\O(\sqrt{n} \ell_j)$, where $\ell_j$ is the total size of all sublists
$\L'(u_i^j)$.  When we sum this complexity for all vertices $u$, we get the total running time
$\O(\sqrt{n} \ell)$.\qed
\end{proof}

\section{Planar Graphs} \label{sec:planar_graphs}

In this section, we describe how to solve \lgi on planar graphs.

For the purpose of this section, we need to consider a more general definition of a graph. We work
with multigraphs and we admit pendant edges with free ends (which are edges attached to single
vertices). Also, each edge $uv$ gives rise to two incident darts,\footnote{In the standard
definition of graphs, the primary objects are vertices and the secondary objects are edges. The
definition via darts, from algebraic and topological graph theory, makes edges (or more precisely
their halves) the primary objects, while the vertices are secondary objects. It is important because
we need to distinguish between an isomorphism which maps an edge and which also reflects it.}
one attached to $u$, the other to $v$.  Every isomorphism maps vertices and darts while preserving
incidencies.  We consider the problem \lgi with lists on both vertices and darts.

\heading{3-connected Planar Graphs.}
We have a unique embedding into the sphere (up to the reflection). This embeddings can be described
in the language of flags, which are pairs $(d,f)$ where $d$ is a dart and $f$ is an incident face.
Every automorphism of $G$ corresponds either to a direct map automorphism, or to a indirect map
automorphism (composed with a reflection). In particular, $\Aut(G) \cong \Aut(\calM)$ acts
semiregularly on the set of flags of $\calM$.  See~\cite{knz} for more details and references.
Therefore, if the images of two consecutive darts in the rotational scheme are set, the entire
mapping is determined and we just need to check whether it is an isomorphism.

\begin{lemma} \label{lem:planar_3conn}
The problem \lgi (with lists on both vertices and darts) can be solved for 3-connected planar graphs
in time $\O(\ell)$.
\end{lemma}

\begin{proof}
We start by computing embeddings of both $G$ and $H$, in time $\O(n)$. It remains to decide whether
there exists a list-compatible isomorphism which has to be a map isomorphism. By Euler Theorem, we
know that the average degree is less than six. Consider all vertices of degree at most 5, let $u$ be
such a vertex with a smallest list, and let $k = |\L(u)|$. We have $\ell = \Omega(kn)$ and we show
that we can decide existence of a list-compatible isomorphism in time $\O(kn)$.

We test all possible mappings $\pi : G \to H$ having $\pi(u) \in \L(u)$. For each, we have at
most 10 possible ways how to extend this mapping on the neighbors of $u$, and the rest of the
mapping is uniquely determined by the embeddings and can be computed in time $\O(n)$. In the end, we
test whether the constructed mapping $\pi$ is an isomorphism and whether it is list-compatible.\qed
\end{proof}

\heading{3-connected Reduction.}
A seminal paper by Trakhtenbrot~\cite{trakhtenbrot} introduced reduction which decomposes a graph
into its \emph{3-connected components}. This idea was further extended
in~\cite{tutte_connectivity,quadratic_isomorphism_planar,hopcroft_tarjan_dividing,cunnigham_edmonds,walsh}.
We use an augmentation described in~\cite{fkkn,fkkn16,knz} which behaves well with respect to
automorphism groups. 

The reduction is constructed by replacing atoms by colored possibly directed edges. \emph{Atoms} are
subgraphs of the following three types (for precise definitions, see~\cite{fkkn16}):
\begin{packed_itemize}
\item \emph{Block atom.} Either a pendant star, or a pendant block with attached single pendant
edges.
\item \emph{Proper atom.} Inclusion minimal subgraphs separated by a 2-cut.
\item \emph{Dipoles.} They are two vertices together with all (at least two) parallel edges between
them.
\end{packed_itemize}
Further, each atom $A$ has the \emph{boundary} $\bo A$ (of size at most 2) and the \emph{interior}
$\int A$. A graph is called \emph{essentially 3-connected} if it is a 3-connected graph with
attached single pendant edges attached. Similarly, a graph is called \emph{essentially a cycle} if
it is a cycle with attached single pendant edges.  It follows from~\cite{fkkn16} that each block
atom is either a star, or essentially a cycle, or essentially 3-connected, or $K_2$ with a single
pendant edge attached. For a proper atom $A$ with $\bo A = \{u,v\}$, we denote by $A^+$ the graph
with the added edge $uv$. The graph $A^+$ is always either essentially a cycle, or essentially
3-connected.

A proper atom or a dipole $A$ is called \emph{symmetric} if there exists an automorphism in
$\Aut(A)$ exchanging $\bo A$, and \emph{asymmetric} otherwise. Every block atom is symmetric by the
definition. The reduction is done by finding all atoms in $G$ (by~\cite{fkkn16}, they have disjoint
interiors) and replacing their interiors by edges. Further, we color these edges to code isomorphism
types of atoms, and we use directed edges for asymmetric atoms. Block atoms are replaced by pendant
edges with free ends.

\begin{figure}[t!]
\centering
\includegraphics[scale=1]{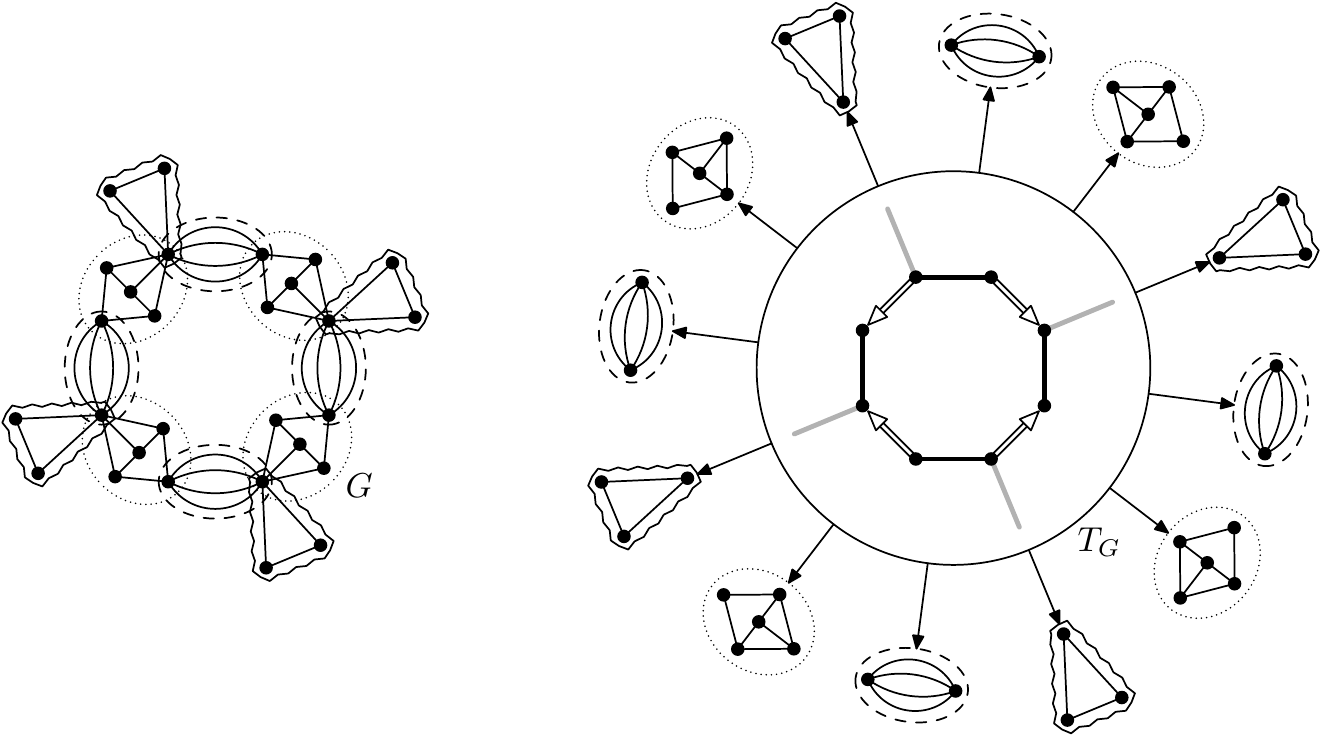}
\caption{A graph $G$ together with its reduction tree $T_G$.}
\label{fig:reduction_tree}
\end{figure}

We repeat this reductions over and over, which gives a sequence of graphs $G = G_0, \dots, G_r$
where $G_r$ is called \emph{primitive} and contains no atoms. By~\cite{fkkn16}, it is either
essentially 3-connected, essentially a cycle, $K_2$ possibly with attached single pendant edges, or
$K_1$ with an attached single pendant edge with a free end. Further, this reduction process can be
encoded by the \emph{reduction tree} $T_G$; see Fig.~\ref{fig:reduction_tree} for an example.
It is a rooted tree, where each node is labeled by a graph. The root of $T_G$ is the primitive graph
$G_r$. The other nodes correspond to atoms obtained in the reductions. When the interior of an atom
$A$ is replaced by an edge $e$, we attach the node representing to $A$ to the edge $e$.

It easily follows that the reduction tree is unique and canonical. Further each automorphism $\pi$
of $G$ induces automorphisms of $G_1,\dots,G_r$ by permuting edges exactly as atoms. Therefore, 
it induces an automorphism $\pi'$ of $T_G$ which permutes the nodes of isomorphic graphs, and when
it maps a colored edge $e$ to a colored edge $e'$, it maps the subtree attached to $e$ to the
isomorphic subtree attached to $e'$. And every automorphism of $G$ can be constructed in this way,
from the root of $T_G$ to the bottom. We can use this to solve \lgi.

\begin{theorem} \label{thm:3conn_reduction}
Let $\calC$ be a class of graphs closed under contractions and removing vertices. Suppose that \lgi
with lists on both vertices and darts can be solved for 3-connected graphs in $\calC$ in time
$\varphi(n,m,\ell)$. We can solve \lgi on $\calC$ in time $\O(\sqrt m \ell + m +
\varphi(n,m,\ell))$.
\end{theorem}

\begin{proof}
We compute reduction trees $T_G$ and $T_H$ for both $G$ and $H$ in time $\O(n+m)$.  We apply the
idea of Theorem~\ref{thm:trees} to test list-compatible isomorphism of $T_G$ and $T_H$. We compute
the lists $\L(N)$ for the nodes $N$ of $T_G$, from the bottom to the root. A node $M \in \L(N)$ if
there exists a list-compatible isomorphism from $N$ to $M$ mapping $\bo N$ to $\bo M$ and there
exists list-compatible isomorphism between attached subtrees. (Further, if $|\bo N| = |\bo M| = 2$,
we remember in $\L(N)$ which of both possible mappings of $\bo N$ to $\bo M$ can be extended as
list-compatible isomorphisms.)

Suppose that $N$ has the children $N_1,\dots,N_k$ with computed lists and $M$ has the children
$M_1,\dots,M_k$.  There exists a list-compatible isomorphism mapping the subtree of $N_i$ to the
subtree of $M_j$, if and only if $M_j \in \L(N_i)$. The difference from Theorem~\ref{thm:trees} is
these subtrees have to be compatible with a list-isomorphism from $N$ to $M$; so it depends on the
structure of the nodes $N$ and $M$.

There, we compute $\L(N)$ differently according to the type of $N$:
\begin{packed_itemize}
\item \emph{Star block atoms or dipoles.} For star block atoms, similarly as in
Theorem~\ref{thm:trees}, we construct a bipartite graph between $N_1,\dots,N_k$ and $M_1,\dots,M_k$
and test existence of a perfect matching using Lemma~\ref{lem:bipmatch}. For dipoles, we test two
possible isomorphisms, construct two bipartite graph and test existence of perfect matchings.
\item \emph{Non-star block or proper atoms.} We modify the lists of $\bo N$ to the vertices of $\bo
M$ only. (When they are proper atoms, we run this in two different ways.) We encode the lists
$\L(N_1),\dots,\L(N_k)$ by lists on the corresponding darts of $N$ (depending on which of two
possible list-isomorphisms of $\bo N_i$ are possible), and we remove single pendant edges, and
intersect their lists with the lists of the incident vertices. For a proper atom, we further
consider $N^+$ and $M^+$ with added edges $e$ and $f$ such that $\L(e) = \{f\}$. If the nodes are
$K_2$ or cycles, and we can test existence of a list-compatible isomorphism using
Lemma~\ref{lem:cycles}. If both are 3-connected, we can test it by our assumption in time
$\varphi(n',m',\ell')$. If this list-compatible isomorphism exists, we add $M$ to $\L(N)$.
\item \emph{The root primitive graphs.} We use the same approach as above, ignoring the part
about $\bo N$ and $\bo M$.
\end{packed_itemize}
A list-compatible isomorphism from $G$ to $H$ exists, if and only if $M \in \L(N)$ for the root
nodes $N$ and $M$ of $T_G$ and $T_H$.

The correctness of the algorithm can be argued from the fact that all automorphisms are captured by
the reduction trees~\cite{fkkn16}, inductively from the top to the bottom as in
Theorem~\ref{thm:trees}.  It remains to discuss the running time. The reduction trees can be
computed in linear time~\cite{hopcroft_tarjan_dividing}. When computing $\L(N)$, we first consider
the lists of all vertices and edges of $N$. A node $M$ is a candidate for $\L(N)$, if every vertex
and every edge of $N$ has a vertex/edge of $M$ in its list. Therefore, we can find all these
candidate nodes by iterating these lists, in linear time with respect to their total size. Let $M$
be one of them, and let $n' = |V(N)|$, $m' = |E(N)|$ and $\ell'$ be the total size of lists of the
vertices and edges of $N$ when restricted only to the vertices and edges of $M$. Either we construct
a bipartite graph and test existence of a perfect matching in time $\O(\sqrt m' \ell')$, or we test
existence of a list-compatible isomorphism in time $\varphi(n',m',\ell')$. The total running time
spent on the tree is $\O(\ell)$, the total running time spent testing perfect matchings is $\O(\sqrt
m \ell)$, and the total running time testing list-compatible isomorphisms of 3-connected graphs is
$\O(\varphi(n,m,\ell))$.\qed
\end{proof} 

\heading{General Planar Graphs.}
By putting both results together, we get the following result:

\begin{theorem} \label{thm:planar}
The problem \lgi can be solved for planar graphs in time $\O(\sqrt n \ell)$.
\end{theorem}

\begin{proof}
If $G$ and $H$ are connected, we use Theorem~\ref{thm:3conn_reduction}.  By
Lemma~\ref{lem:planar_3conn}, the function $\varphi(n,m,\ell)$ is $\O(\ell)$. If $G$ and $H$ are
disconnected, we apply Lemma~\ref{lem:disconnected} on all connected components of $G$ and $H$, and
by analysing the proof, the total running time is $\O(\sqrt n \ell)$.\qed
\end{proof}

\section{Interval, Permutation and Circle Graphs} \label{sec:intersection}

In this section, we prove that the standard algorithms solving \gi on interval, circle and
permutation graphs can be modified to solve \lgi on them. The key idea is that the structure of
these graph classes can be captured by graph-labeled trees which are unique up to isomorphism and
which capture the structure of all automorphisms; see~\cite{kz,kz15} and the references therein.

For interval graphs, we use MPQ-tree. For circle graphs, we use split trees. For permutation graphs,
we use modular trees. On these trees, we apply bottom-up procedure similarly as in the proof of
Theorem~\ref{thm:trees}. The key difference is that nodes correspond to either prime, or degenerate
graphs. Degenerate graphs are simpler and lead to perfect matchings in bipartite graphs.  Prime
graphs have a small number of automorphisms~\cite{kz,kz15}, so all of them can be tested.

\subsection{Interval Graphs}

To each interval graph $G$, a unique MPQ-tree $T_G$ is assigned. Two interval graphs $G$ and $H$ are
isomorphic if and only if $T_G$ and $T_H$ are equivalent, and these trees capture all isomorphisms.
Therefore, we apply a bottom-up proceduce to test \lgi for MPQ-trees, similarly as in
Theorem~\ref{thm:trees}.

\heading{MPQ-trees.} Booth and Lueker~\cite{PQ_trees} invented a data structure called a
\emph{PQ-tree} which capture the structure of an interval graph.  We use \emph{modified PQ-trees}
(MPQ-trees) due to Korte and M\"ohring~\cite{incremental_linear_int_recognition}.  Let $G$ be an
interval graph.  A rooted tree $T$ is an MPQ-tree if the following holds.  It has two types of inner
nodes: \emph{P-nodes} and \emph{Q-nodes}. For every inner node, its children are ordered from left
to right. Each P-node has at least two children and each Q-node at least three.  The leaves of $T$
correspond one-to-one to the maximal cliques in $G$.

\begin{figure}[t]
\centering
\includegraphics[scale=1]{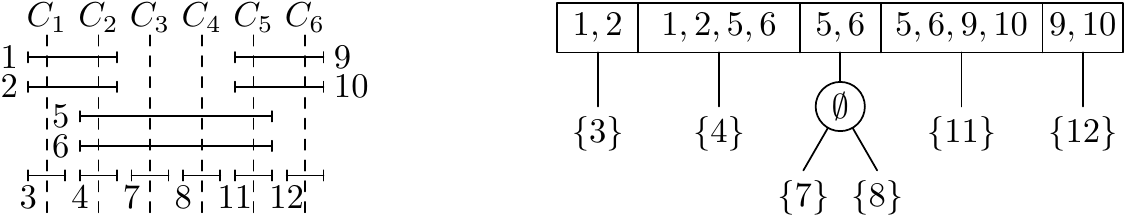}
\caption{An ordering of the maximal cliques, and the corresponding MPQ-tree. The P-nodes
are denoted by circles, the Q-nodes by rectangles. There are four equivalent MPQ-trees.}
\label{fig:mpq_tree}
\end{figure}

Two MPQ-trees are \emph{equivalent} if one can be obtained from the other by a sequence of two
\emph{equivalence transformations}: (i) an arbitrary permutation of the order of the children of a
P-node, and (ii) the reversal of the order of the children of a Q-node.  Booth and
Lueker~\cite{PQ_trees} proved the existence and uniqueness of PQ-trees (up to equivalence
transformations); see Fig.~\ref{fig:mpq_tree}.

We assign subsets of $V(G)$, called \emph{sections}, to the
nodes of $T$; see Fig.~\ref{fig:mpq_tree}. The leaves and the P-nodes have
each assigned exactly one section while the Q-nodes have one section per child.
We assign these sections as follows:
\begin{itemize}
\item For a leaf $L$, the section $\sec(L)$ contains those vertices that are only
in the maximal clique represented by $L$, and no other maximal clique.
\item For a P-node $P$, the section $\sec(P)$ contains those vertices that are in
all maximal cliques of the subtree of $P$, and no other maximal clique.
\item For a Q-node $Q$ and its children $T_1, \dots, T_n$, the section $\sec_i(Q)$
contains those vertices that are in the maximal cliques represented by the
leaves of the subtree of $T_i$ and also some other $T_j$, but not in any other
maximal clique outside the subtree of $Q$. We put $\sec(Q) = \sec_1(Q) \cup
\cdots \cup \sec_n(Q)$.
\end{itemize}
Each vertex appears in sections of exactly one node and in the case of a Q-node in consecutive
sections. Two vertices are in the same sections if and only if they belong to precisely the same
maximal cliques. Figure~\ref{fig:mpq_tree} shows an example. MPQ-tree can be constructed in
time~\cite{incremental_linear_int_recognition}. 

\heading{Testing \lgi.}
Let $G$ and $H$ be two isomorphic interval graphs. From~\cite[Lemma 4.3]{kz15}, it follows that
$T_G$ and $T_H$ are equivalent, and every isomorphism $\pi : G \to H$ is obtained by an equivalence
transformation of $T_G$ and some permutation of the vertices in identical sections.  Now, we are
ready to show \lgi can be solved on interval graphs in time $\calO(\sqrt{n}\ell + m)$:

\begin{theorem} \label{thm:int}
The problem \lgi can be solved for interval graphs in time $\calO(\sqrt{n}\ell + m)$.
\end{theorem}

\begin{proof}
We proceed similarly as in Theorem~\ref{thm:trees}.  We compute MPQ-trees representing $T_G$
and $T_H$ representing the graphs $G$ and $H$ in linear
time~\cite{incremental_linear_int_recognition}. Then we compute lists $\L(N)$ for every node $N$ of
$T_G$ from the bottom. We distinguish three types of nodes.
\begin{itemize}
\item \emph{Leaf nodes.} Let $L_G$ be a leaf node in $T_G$ and let $L_H$ be a leaf node in $T_H$.
Then $L_H \in \L(L_G)$ if there exists a list-compatible isomorphism between the induced complete
subgraphs $G[\sec(L_G)]$ and $H[\sec(L_H)]$.
\item \emph{P-nodes.} Let $N$ and $M$ be P-nodes of $T_G$ and $T_H$, respectively. We want to decide
whether $M \in \L(N)$. Let $N_1,\dots,N_k$ be the children of $N$ and let
$M_1,\dots,M_k$ be the children of $M_k$. We construct a bipartite graph
similarly as in Theorem~\ref{thm:trees}.  Then $M \in \L(N)$ if there exists a
perfect matching in the bipartite graph and a perfect matching between the lists of 
$G[\sec(N)]$ and $H[\sec(M)]$ (which are complete graphs).
\item \emph{Q-nodes.} Let $N$ and $M$ be Q-nodes of $T_G$ and $T_H$ and let $N_1,\dots,N_k$ and
$M_1,\dots,M_k$ be their children. Here we have at most two possible
isomorphisms. In particular, an isomorphism can either map the subtree of $N_i$
on the subtree of $M_i$, or in the reversed order, and we can test for both possibilities whether
the lists $\L(N_i)$ are compatible. Moreover, we consider all sets of intervals belonging to
exactly the same sections of the Q-node, and we test by perfect matchings between pairs of them
whether there exists a list-compatible isomorphism between them.
\end{itemize}
The MPQ-trees have $\O(n)$ nodes and $\O(n)$ intervals in their sections. For leaf nodes and
P-nodes, the analysis is exactly the same as in the proof of Theorem~\ref{thm:trees}. For Q-nodes,
we just test two possible mappings and bipartite matchings for sections. We get the total running
time $\O(\sqrt n \ell + m)$.\qed
\end{proof}

\subsection{Permutation Graphs}

\begin{figure}[b]
\centering
\includegraphics[scale=1]{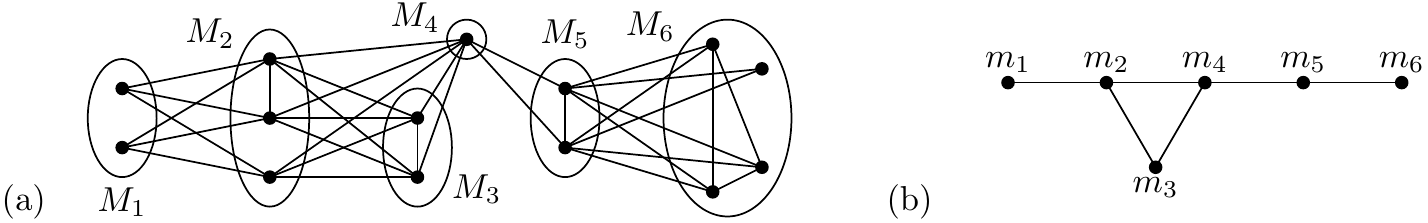}
\caption{(a) A graph $G$ with a modular partition $\calP$. (b) The quotient graph $G/\calP$ is prime.}
\label{fig:modules}
\end{figure}

A \emph{module} $M$ of a graph $G$ is a set of vertices such that each $x \in V(G) \setminus M$ is
either adjacent to all vertices in $M$, or to none of them.  See Fig.~\ref{fig:modules}a for
examples. A module $M$ is called \emph{trivial} if $M=V(G)$ or $|M|=1$, and \emph{non-trivial}
otherwise. If $M$ and $M'$ are two disjoint modules, then either the edges between $M$ and $M'$
form the complete bipartite graph, or there are no edges at all; see Fig.~\ref{fig:modules}a.  In
the former case, $M$ and $M'$ are called \emph{adjacent}, otherwise they are \emph{non-adjacent}.

Let $\calP = \{M_1, \dots, M_k\}$ be a \emph{modular partition} of $V(G)$, i.e., each $M_i$ is a
module of $G$, $M_i \cap M_j = \emptyset$ for every $i\neq j$, and $M_1 \cup \cdots \cup M_k =
V(G)$. We define the \emph{quotient graph} $G/\calP$ with the vertices $m_1,\dots,m_k$ corresponding
to $M_1,\dots,M_k$ where $m_im_j \in E(G/\calP)$ if and only if $M_i$ and $M_j$ are adjacent.  In
other words, the quotient graph is obtained by contracting each module $M_i$ into the single vertex
$m_i$; see Fig.~\ref{fig:modules}b.

\heading{Modular Decomposition.}
To decompose $G$, we find some modular partition $\calP = \{M_1, \dots, M_k\}$, compute $G / \calP$
and recursively decompose $G / \calP$ and each $G[M_i]$. The recursive process terminates on
\emph{prime graphs} which are graphs containing only trivial modules.  There might be many such
decompositions for different choices of $\calP$ in each step.  In 1960s,
Gallai~\cite{gallai1967transitiv} described the \emph{modular decomposition} in which special
modular partitions are chosen and which encodes all other decompositions.

\begin{figure}[t]
\centering
\includegraphics[scale=1]{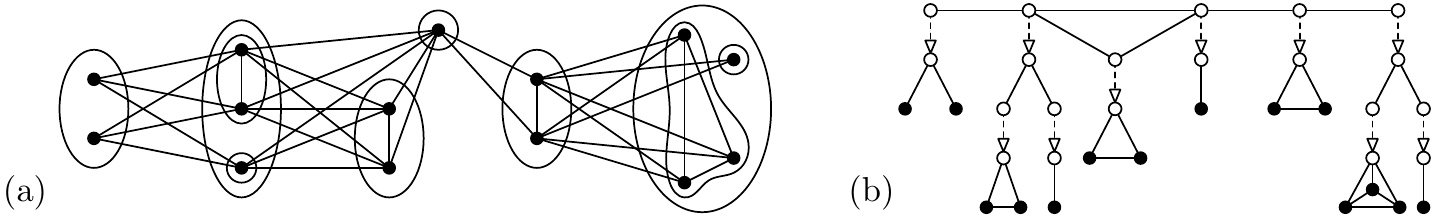}
\caption{(a) The graph $G$ from Fig.~\ref{fig:modules} with the modular partitions used in the
modular decomposition. (b) The modular tree $T$ of $G$, the marker vertices are white, the tree
edges are dashed.}
\label{fig:modular_tree}
\end{figure}

The key is the following observation. Let $M$ be a module of $G$ and let $M' \subseteq M$. Then $M'$
is a module of $G$ if and only if it is a module of $G[M]$.  A graph $G$ is called \emph{degenerate}
if it is $K_n$ or $\overline{K}_n$.	We construct the modular decomposition of a graph $G$ in the
following way, see Fig.~\ref{fig:modular_tree}a for an example:
\begin{itemize}
\item If $G$ is a prime or a degenerate graph, then we terminate the modular decomposition on $G$.
We stop on degenerate graphs since every subset of vertices forms a module, so it is not useful to
further decompose them.
\item Let $G$ and $\overline{G}$ be connected graphs.  Gallai~\cite{gallai1967transitiv} shows that
the inclusion maximal proper subsets of $V(G)$ which are modules form a modular partition $\calP$ of
$V(G)$, and the quotient graph $G/\calP$ is a prime graph; see Fig.~\ref{fig:modules}. We
recursively decompose $G[M]$ for each $M \in \calP$.
\item If $G$ is disconnected and $\overline{G}$ is connected, then every union of connected
components is a module. Therefore the connected components form a modular partition $\calP$ of
$V(G)$, and the quotient graph $G/\calP$ is an independent set. We recursively decompose $G[M]$ for
each $M \in \calP$.
\item If $\overline{G}$ is disconnected and $G$ is connected, then the modular decomposition is
defined in the same way on the connected components of $\overline{G}$.  They form a modular
partition $\calP$ and the quotient graph $G/\calP$ is a complete graph. We recursively decompose
$G[M]$ for each $M \in \calP$.
\end{itemize}

\heading{Modular Tree.}
We encode the modular decomposition by the \emph{modular tree} $T$.  The modular tree $T$ is a graph
with two types of vertices (normal and \emph{marker vertices}) and two types of edges (normal and
\emph{directed tree edges}). The directed tree edges connect the prime and degenerate graphs
encountered in the modular decomposition (as quotients and terminal graphs) into a rooted tree. 

We give a recursive definition.  Every modular tree has an induced subgraph called \emph{root node}.
If $G$ is a prime or a degenerate graph, we define $T = G$ and its root node equals $T$.  Otherwise,
let $\calP = \{M_1,\dots,M_k\}$ be the used modular partition of $G$ and let $T_1,\dots,T_k$ be the
modular trees corresponding to $G[M_1],\dots,G[M_k]$. The modular tree $T$ is the disjoint union of
$T_1,\dots,T_k$ and of $G / \calP$ with the marker vertices $m_1,\dots,m_k$. To every graph $T_i$,
we add a new marker vertex $m'_i$ such that $m'_i$ is adjacent exactly to the vertices of the root
node of $T_i$. We further add a tree edge oriented from $m_i$ to $m'_i$.  For an example, see
Fig.~\ref{fig:modular_tree}b.

The modular tree of $G$ is unique.  The graphs encountered in the modular decomposition are called
\emph{nodes of $T$}, or alternatively root nodes of some modular trees in the construction of $T$.
For a node $N$, its subtree is the modular tree which has $N$ as the root node. \emph{Leaf nodes}
correspond to the terminal graphs in the modular decomposition, and \emph{inner nodes} are the
quotients in the modular decomposition. All vertices of $G$ are in leaf nodes and all marker
vertices correspond to modules of $G$. \emph{All inner nodes consist of marker vertices}.

\heading{Testing \lgi.}
Now, we are ready to show that the problem \lgi can be decided in $\calO(\sqrt{n}\ell + m)$ time for
permutation graphs.

\begin{theorem} \label{thm:perm}
The problem \lgi can be solved for permutation graphs in time $\calO(\sqrt{n}\ell + m)$.
\end{theorem}

\begin{proof}
For input graph $G$ and $H$, we first compute the modular trees $T_G$ and
$T_H$, respectively, in time $\O(n+m)$~\cite{mcconnell_spinrad}. We again apply
the idea of Theorem~\ref{thm:trees}.  We compute the list $\L(N)$ for every
node $N$ of $T_G$. Note that all inner nodes consist only of marker vertices
which have no lists. Therefore, we first compute $\L(L)$, for every leaf node.
A leaf node $K$ is in $\L(L)$ if every non-marker vertex of $L$ has a
non-marker vertex of $K$ in its list. These candidate nodes for $\L(L)$ can be
found in linear time in the total size of lists by iterating through the lists
of vertices of $L$.

Suppose that a node $N$ has the children $N_1,\dots,N_k$ with computed lists $\L(N_1),\dots,\L(N_k)$
and $M$ has the children $M_1,\dots,M_k$. There exist a list-compatible isomorphism mapping the
subtree of $N_i$ to the subtree of $M_j$ if and only if $M_j \in \L(N_i)$. Moreover these subtrees
have to be compatible with a list isomorphism from $N$ to $M$. We compute $\L(N)$ according to the
type of $N$.
\begin{packed_itemize}
\item \emph{Degenerate nodes.} For degenerate nodes, we proceed similarly as for trees in
Theorem~\ref{thm:trees}. We construct a bipartite graph between the nodes nodes $N_1,\dots,N_k$ and
$M_1,\dots,M_k$ and test for a perfect matching using Lemma~\ref{lem:bipmatch}.
\item \emph{Prime nodes.} For prime nodes, there are at most four possible isomorphisms mapping $N$
to $M$~\cite[Lemma 6.6]{kz15}. We test for these four possible isomorphisms $\pi$ whether $\pi(M_i)
\in \L(M_i)$ for every $M_i$.
\end{packed_itemize}
A list compatible isomorphism exists if $M \in \L(N)$, for the root nodes $N$ and $M$ of $T_G$ and
$T_H$.  The correctness of the algorithm follows from the fact that all
automorphisms of a permutation
graph are captured by the modular tree~\cite{kz15}. A similar argument as in the proofs of
Theorems~\ref{thm:trees} and~\ref{thm:int} gives the running time.\qed
\end{proof}

\subsection{Circle Graphs}

For a given circle graph, we define the split tree which captures its automorphism group.  A
\emph{split} is a partition $(A,B,A',B')$ of $V(G)$ such that:
\begin{itemize}
\item For every $a \in A$ and $b \in B$, we have $ab \in E(G)$.
\item There are no edges between $A'$ and $B \cup B'$, and between $B'$ and $A \cup A'$.
\item Both sides have at least two vertices: $|A \cup A'| \ge 2$ and $|B \cup B'| \ge 2$.
\end{itemize}

\begin{figure}[b]
\centering
\includegraphics[scale=1]{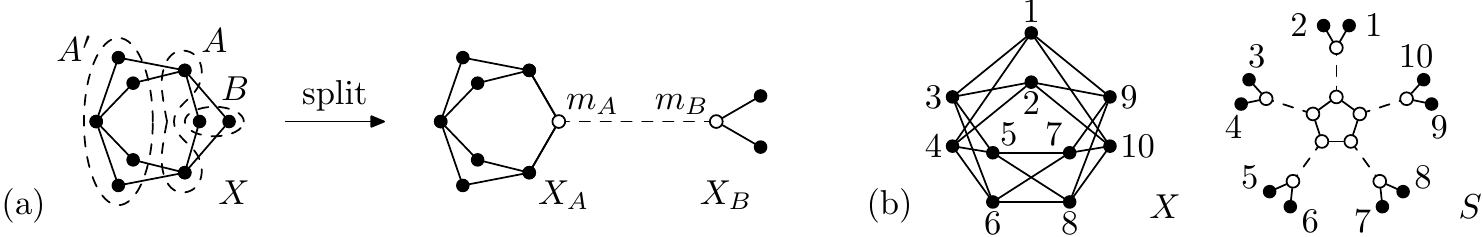}
\caption{(a) An example of a split of the graph $G$. The marker vertices are depicted in white.  The
tree edge is depicted by a dashed line.  (b) The split tree $S$ of the graph $G$. We have that
$\Aut(S) \cong \cyc_2^5 \rtimes \dih_5$.}
\label{fig:split_graph_split-tree}
\end{figure}

The split decomposition of $G$ is constructed by taking a split of $G$ and replacing $G$ by the
graphs $G_A$ and $G_B$ defined as follows. The graph $G_A$ is created from $G[A \cup A']$ together
with a new \emph{marker vertex} $m_A$ adjacent exactly to the vertices in $A$.  The graph $G_B$ is
defined analogously for $B$, $B'$ and $m_B$; see Fig.~\ref{fig:split_graph_split-tree}a. The
decomposition is then applied recursively on $G_A$ and $G_B$. Graphs containing no splits are called
\emph{prime graphs}. We stop the split decomposition also on \emph{degenerate graphs} which are
complete graphs $K_n$ and stars $K_{1,n}$. A split decomposition is called \emph{minimal} if it is
constructed by the least number of splits. Cunningham~\cite{cunningham} proved that the minimal
split decomposition of a connected graph is unique.

\heading{Split tree.}
The \emph{split tree} $T$ representing a graph $G$ encodes the minimal split decomposition. A split
tree is a graph with two types of vertices (normal and marker vertices) and two types of edges
(normal and tree edges). We initially put $T = G$ and modify it according to the minimal split
decomposition. If the minimal decomposition contains a split $(A, B, A', B')$ in $G$, then we
replace $G$ in $T$ by the graphs $G_A$ and $G_B$, and connect the marker vertices $m_A$ and $m_B$ by
a \emph{tree edge} (see Fig.~\ref{fig:split_graph_split-tree}a). We repeat this recursively on $G_A$
and $G_B$; see Fig.~\ref{fig:split_graph_split-tree}b. Each prime and degenerate graph is a
\emph{node} of the split tree. A node that is incident with exactly one tree edge is called a
\emph{leaf node}.

Since the minimal split decomposition is unique, we also have that the split tree is unique.
Further, each automorphism $\pi$ of $G$ induces an automorphism $\pi'$ of the split tree $T$
representing $G$. Similarly as for trees, there exists a \emph{center} of $T$ which is either a tree
edge, or a prime or degenerate node. The automorphism $\pi'$ preserves the center, so we can regard
$T$ as rooted by the center. Every automorphism of $G$ can be reconstructed from the root of $T$ to
the bottom.

\heading{Testing \lgi.}
Next, we show that the problem \lgi can be solved on circle graphs in time $\calO(\sqrt{n}\ell +
m\cdot \alpha(m)$.

\begin{theorem} \label{thm:circle}
The problem \lgi can be solved for circle graphs in time $\calO(\sqrt{n}\ell + m\cdot \alpha(m))$,
where $\alpha$ is the inverse Ackermann function.
\end{theorem}

\begin{proof}
For input graph $G$ and $H$, we first compute the split trees $T_G$ and $T_H$, in time
$\O((n+m)\cdot \alpha(n+m))$~\cite{split_trees_linear}. We assume that the trees $T_G$ and $T_H$ are
rooted and we can also assume that the roots are prime or degenerate nodes. We again apply the idea
of Theorem~\ref{thm:trees}.

We compute the list $\L(N)$ for every node $N$ of $T_G$. Let $M$ be a leaf node of $T_H$ and let
$m_N \in V(N)$ and $m_M \in V(M)$ be the marker vertices incident to a tree edge closer to the root.
Then $M$ is in $\L(N)$ if there is a list-compatible isomorphism from $N$ to $M$ which maps $m_N$
to $m_M$.

Suppose that a node $N$ has the children $N_1,\dots,N_k$ with computed lists $\L(N_1),\dots,\L(N_k)$
and $M$ has the children $M_1,\dots,M_k$. There exist a list-compatible isomorphism mapping the
subtree of $N_i$ to the subtree of $M_j$ if and only if $M_j \in \L(N_i)$. Moreover these subtrees
have to be compatible with an isomorphism from $N$ to $M$. We compute $\L(N)$ according to the
type of $N$.

\begin{packed_itemize}
\item \emph{Degenerate nodes.} For degenerate nodes, we proceed similarly as for trees in
Theorem~\ref{thm:trees}. We construct a bipartite graph between the nodes nodes $N_1,\dots,N_k$ and
$M_1,\dots,M_k$ and test for a perfect matching using Lemma~\ref{lem:bipmatch}.
\item \emph{Prime nodes.} For prime nodes, there are at most four possible isomorphisms mapping
$m_N$ to $m_M$~\cite[Lemma 5.6]{kz}. We test those four possible isomorphisms, construct four bipartite graphs
and test existence of perfect matchings.
\item \emph{The root node.} If it is degenerate, we proceed as above. If it is prime, then its
automorphism groups is a subgroup of a dihedral group~\cite[Lemma 5.5]{kz}; essentially it behaves
as a cycle. Therefore, we approach it similarly as in Lemma~\ref{lem:cycles}.
\end{packed_itemize}
A list compatible isomorphism exists if $M \in \L(N)$, for the root nodes $N$
and $M$ of $T_G$ and $T_H$.

The correctness of the algorithm follows from the fact that all automorphisms of a circle graph are
captured by the split tree~\cite{kz}. The running time can be argued as in Theorems~\ref{thm:trees},
\ref{thm:int}, and~\ref{thm:perm}.\qed
\end{proof}

\section{Bounded Genus Graphs} \label{sec:bounded_genus}

In this section, we describe an FPT algorithm solving \lgi when parameterized by the Euler genus
$g$. We modify the recent paper of Kawarabayashi~\cite{kawarabayashi} solving graph isomorphism in
linear time for a fixed genus $g$. The harder part of this paper are structural results, described
below, which transfer to list-compatible isomorphisms without any change. Using these structural
results, we can build our algorithm.

\begin{theorem} \label{thm:bounded_genus}
For every integer $g$, the problem \lgi can be solved on graphs of Euler genus at most $g$ in time
$\O(\sqrt n \ell)$.
\end{theorem}

\begin{proof}
See~\cite[p.~14]{kawarabayashi} for overview of the main steps. We show that these steps can be
modified to deal with lists. We prove this result by induction on $g$, where the base case for $g=0$
is Theorem~\ref{thm:planar}. Next, we assume that both graphs $G$ and $H$ are 3-connected, otherwise
we apply Theorem~\ref{thm:3conn_reduction}.  By~\cite[Theorem 1.2]{kawarabayashi}, if $G$ and $H$
have no polyhedral embeddings, then the face-width is at most two.
	 
\emph{Case 1: $G$ and $H$ have polyhedral embeddings.} Following~\cite[Theorem 1.2]{kawarabayashi},
we have at most $f(g)$ possible embeddings of $G$ and $H$. We choose one embedding of $G$ and we
test all embeddings of $H$. It is known that the average degree is $\O(g)$.  Therefore, we can apply
the same idea as in the proof of Lemma~\ref{lem:planar_3conn} and test isomorphism of all these
embeddings in time $\O(\ell)$. 

\emph{Case 2: $G$ and $H$ have no polyhedral embedding, but have embeddings of face-width exactly two.}
Then we split $G$ into a pair of graphs $(G',L)$. The graph $L$
are called \emph{cylinders} and the graph $G'$ correspond to the remainder of $G$. The following
properties hold~\cite[p.~5]{kawarabayashi}:
\begin{packed_itemize}
\item We have $G = G' \cup L$ and for $\bo L = V(G' \cap L)$, we have $|\bo L| = 4$.
\item The graph $G'$ can be embedded to a surface of genus at most $g-1$, and $L$ is
planar~\cite[p.~4]{kawarabayashi}.
\item This pair $(G',L)$ is canonical, i.e., every isomorphism from $G$ to $H$ maps $(G',L)$ to another
pair $(H',L')$ in $H$.
\end{packed_itemize}
It is proved~\cite[Theorem 5.1]{kawarabayashi} that there exists some function $q'(g)$ bounding the
number of these pairs both in $G$ and $H$, and can be found in time $\O(n)$. We fix a pair $(G',L)$
in $G$ and iterate over all pairs $(H'_i,L'_i)$ in $H$.  Following~\cite[p.~36]{kawarabayashi}, we
get that $G \cong H$, if and only if there exists a pair $(H'_i,L'_i)$ in $H$ such that $G' \cong
H'_i$, $L \cong L'_i$, and $G' \cap L$ is mapped to $H'_i \cap L'_i$. To test this, we run at most
$2q'(g)$ instances of \lgi on smaller graphs with modified lists.

Suppose that we want to test whether $G' \cong H'_i$ and $L \cong L'_i$. First, we modify the lists:
for $u \in V(G')$, put $\L'(u) = \L(u) \cap H'_i$, and for $v \in V(L)$, put $\L'(v) = \L(v) \cap
L'_i$, and similarly for lists of darts. Further, for all vertices $u \in \bo L$ in both $G'$ and
$L$, we put $\L'(u) = \L(u) \cap \bo L$. We test existence of list-compatible isomorphisms from $G'$
to $H'_i$ and from $L$ to $L'_i$. There exists a list-compatible isomorphism from $G$ to $H$, if and
only if these list-compatible isomorphisms exist at least for one pair $(H'_i,L'_i)$.

We note that when $g=2$, a special case is described in~\cite[Theorem 5.3]{kawarabayashi}, which is
slightly easier and can be modified similarly.

\emph{Case 3: $G$ and $H$ have no polyhedral embedding and have only embeddings of face-width one.}
Let $V$ be the set of vertices in $G$ such that for each $u \in V$, there exists a non-contractible
curve passing only through $u$. By~\cite[Lemma 6.3]{kawarabayashi}, $|V| \le q(g)$ for some function
$q$. For $u$, the non-contractible curve divides its edges to two sides, so we can cut $G$ at $u$,
and split the incident edges. We obtain a graph $G'$ which can be embedded to a surface of genus at
most $g-1$.

By~\cite[Lemma 6.3]{kawarabayashi}, we can find all these vertices $V$ and $V'$ in $G$ and $H$ in
time $\O(n)$. We choose $u \in V$ arbitrarily, and we test all possible vertices $v \in V'$.
Let $G'$ be constructed from $G$ by splitting $u$ into new vertices $u'$ and $u''$, and similarly
$H'$ be constructed from $H$ by splitting $v$ into new vertices $v'$ and $v''$. 
In~\cite[p.~36]{kawarabayashi}, it is stated that $G \cong H$, if and only if there exists a choice
of $v \in V'$ such that $G' \cong H'$ and $\{u',u''\}$ is mapped to $\{v',v''\}$.
Therefore, we run at most $q(g)$ instances of \lgi on smaller graphs with modified lists.

If $v \notin L(u)$, clearly a list-compatible isomorphism is not possible for this choice of $v \in
V'$. If $v \in L(u)$, we put $L'(u') = L'(u'') = \{v',v''\}$. Then there exists a list-compatible
isomorphism from $G$ to $H$, if and only if there exists a list-compatible isomorphism from $G'$ to
$H'$.

The correctness of our algorithm follows from~\cite{kawarabayashi}.  It remains to argue the
complexity. Throughout the algorithm, we produce at most $w(g)$ subgraphs of $G$ and $H$, for some
function $w$, for which we test list-compatible isomorphisms. Assuming the induction hypothesis, the
reduction of graphs to 3-connected graphs can be done in time $\O(\sqrt n \ell)$. Case 1 can be
solved in time $\O(\ell)$.  Case 2 can be solved in time $\O(\sqrt n \ell)$. Case 3 can be solved in
time $\O(\sqrt n \ell)$.\qed
\end{proof}

\section{Bounded Treewidth Graphs} \label{sec:bounded_treewidth}

In this section, we prove that \lgi can be solved in \cFPT with respect to the parameter treewidth
$\tw(G)$. Unlike in Sections~\ref{sec:planar_graphs} and~\ref{sec:intersection}, the difficulty
of graph isomorphism on bounded treewidth graphs raises from the fact that
tree decomposition is not uniquely determined.  We follow the approach of
Bodlaender~\cite{iso_xp_treewidth} which describes an \cXP algorithm for \gi of bounded treewidth
graphs, running in time $n^{\O(\tw(G))}$. Then we show that the recent breakthrough by Lokshtanov
et al.~\cite{iso_fpt_treewidth}, giving an \cFPT algorithm for \gi, translates as well.

\begin{definition}
A \emph{tree decomposition} of a graph $G$ is a pair ${\cal T} =
(\{B_i\colon i\in I\},T = (I,F)),$ where $T$ is a rooted tree and $\{B_i\colon
i\in I\}$ is a family of subsets of $V,$ such that
\begin{enumerate}
\item for each $v\in V(G)$ there exists an $i \in I$ such that $v\in B_i$,
\item for each $e\in E(G)$ there exists an $i \in I$ such that $e\subseteq B_i$,
\item for each $v\in V(G), I_v = \{i \in I\colon v\in B_i\}$ induces a subtree of $T.$
\end{enumerate} 
We call the elements $B_i$ the \emph{nodes}, and the elements of the set $F$ the
decomposition edges.
\end{definition}

We define the width of a tree decomposition ${\cal T} = (\{B_i\colon i\in I\},
T)$ as $\max_{i\in I}|B_i|-1$ and the \emph{treewidth} $\tw(G)$ of a graph $G$
as the minimum width of a tree decomposition of the graph $G$. 

\heading{Nice Tree Decompositions.}
It is common to define a {\it nice tree decomposition} of the graph~\cite{kloks}.  We naturally
orient the decomposition edges towards the root and for an oriented decomposition edge $(B_j,B_i)$
from $B_j$ to $B_i$ we call $B_i$ the {\it parent} of $B_j$ and $B_j$ a {\it child} of $B_i$.  If
there is an oriented path from $B_j$ to $B_i$ we say that $B_j$ is a {\it descendant} of $B_i$.

We also adjust a tree decomposition such that for each decomposition edge $(B_i,B_j)$ it holds that
$\big| |B_i|-|B_j| \big| \le 1$ (i.e. it joins nodes that differ in at most one vertex). The
in-degree of each node is at most $2$ and if the in-degree of the node $B_k$ is $2$ then for its
children $B_i,B_j$ holds that $B_i = B_j = B_k$ (i.e. they represent the same vertex set).

We classify the nodes of a nice decomposition into four classes---namely {\it introduce nodes}, {\it
forget nodes}, {\it join nodes} and {\it leaf nodes}. We call the node $B_i$ an introduce node of
the vertex $v$, if it has a single child $B_j$ and $B_i\setminus B_j = \{v\}$. We call the node
$B_i$ a forget node of the vertex $v$, if it has a single child $B_j$ and $B_j\setminus B_i =
\{v\}$. If the node $B_k$ has two children, we call it a join node (of nodes $B_i$ and $B_j$).
Finally we call a node $B_i$ a leaf node, if it has no child.

\heading{Bodlaender's Algorithm.}
A graph $G$ has treewidth at most $k$ if either $|V(G)| \le k$, or there exists a cut set $U
\subseteq V(G)$ such that $|U| \le k$ and each component of $G \setminus U$ together with $U$ has
treewidth at most $k$.  The set $U$ corresponds to a bag in some tree decomposition of $G$.
Bodlaender's algorithm~\cite{iso_xp_treewidth} enumerates all possible cut sets $U$ of size at most
$k$ in $G$ (resp. $H$), we denote these $C_i$ (resp. $D_i$). Furthermore, it enumerates all
connected components of $G\setminus C_i$ as $C_i^j$ (resp.~of $H \setminus D_i$ as $D_i^j$). We
denote by $G[U,W]$ the graph induced by $U \dotcup W$.  The set $W$ is either a connected component
or a collection of connected components. We call $U$ the \emph{border set}.

\begin{lemma}[\cite{ACP87:partialktrees,iso_xp_treewidth}]
\label{lem:partialKTree}
A graph $G[U, W]$ with at least $k$ vertices has a treewidth at most $k$ with the border set $U$ if
and only if there exists a vertex $v\in W$ such that for each connected component $A$ of $G[W
\setminus v]$, there is a $k$-vertex cut $C_s \subseteq U\cup\{v\}$ such that no vertex in $A$ is
adjacent to the (unique) vertex in $(U \cup\{v\})\setminus C_s$, and $G[C_s, A]$ has treewidth at
most $k$.
\end{lemma}

\begin{lemma} \label{lem:xp_treewidth}
The problem \lgi can be solved in \cXP with respect to the parameter treewidth.
\end{lemma}

\begin{proof}
We modify the algorithm of Bodlaender~\cite{iso_xp_treewidth}.  Let $k = \tw(G) = \tw(H)$. We
compute the sets $C_i, C_i^j$ for $G$ and the sets $D_{i'}, D_{i'}^{j'}$ for $H$; there are
$n^{\O(k)}$ pairs $(C_i,C_i^j)$.  The pair $(C_i, C_i^j)$ is \emph{compatible} if $C_i^j$ is a
connected component of $G'\setminus C_i$ for some $G'\subseteq G$ that arises during the recursive
definition of treewidth. Let $f \colon C_i\to D_{i'}$ be an isomorphism. We say that $(C_i,
C_i^j)\equiv_f(D_{i'},D_{i'}^{j'})$ if and only if there exists an isomorphism $\varphi\colon
C_i\cup C_i^j\to D_{i'}\cup D_{i'}^{j'}$ such that $\varphi|_{C_i} = f$.  In other words, $\varphi$
is a partial isomorphism from $G$ to $H$.  The change for \lgi is that we also require that both $f$
and $\varphi$ are list-compatible.

The algorithm resolves $(C_i, C_i^j)\equiv_f(D_{i'},D_{i'}^{j'})$ by the dynamic programming,
according to the size of $D_{i'}^{j'}$. If $|C_i^j| = |D_{i'}^{j'}| \le 1$, we can check it
trivially in time $k^{\O(k)}$. Otherwise, suppose that $|C_i^j| = |D_{i'}^{j'}| > 1$, and let
$m$ be the number of components of $C_i^j$ (and thus $D_{i'}^{j'}$). We test whether $f : C_i \to
D_{i'}$ is a list-compatible isomorphism.  Let $v\in C_i^j$ be a vertex given by
Lemma~\ref{lem:partialKTree} (with $U = C_i$ and $W = C_i^j$) and let $C_s$ be the corresponding
extension of $v$ to a cut set. We compute for all $w \in D_{i'}^{j'} \cap \L(v)$ all connected
components $B_q$. From the dynamic programming, we know for all possible extensions $D'$ of $w$ to a
cut set whether $(C_m, A_p)\equiv_{f'}(D',B_q)$ with $f'(x) = f(x)$ for $x\in C_i$ and $f'(v) = w$.
Finally, we decide whether there exists a perfect matching in the bipartite graph between $(C_m,
A_p)$'s and $(D',B_q)$'s where the edges are according to the equivalence.\qed
\end{proof}

\heading{Reducing The Number of Possible Bags.}
Otachi and Schweitzer~\cite{GIReductionTechniques} proposed the idea of pruning the family of
potential bags which finally led to an \cFPT algorithm~\cite{iso_fpt_treewidth}.  A family
$\mathcal{B}(G)$, whose definition depends on the graph, is called \emph{isomorphism-invariant} if
for an isomorphism $\phi : G \to G'$, we get $\mathcal{B}(G') = \phi(\mathcal{B}(G))$, where
$\phi(\mathcal{B}(G))$ denotes the family $\mathcal{B}(G)$ with all the vertices of $G$ replaced by
their images under $\phi$. 

For a graph $G$, a pair $(A,B)$ with $A \cup B = V$ is called a {\em
separation} if there are no edges between $A\setminus B$ and $B\setminus A$ in
$G$. The order of $(A,B)$ is $|A\cap B|$.  For two vertices $u,v \in V(G)$, by
$\mu(u,v)$ we denote the minimum order of separation $(A,B)$ with $u\in
A\setminus B$ and $v\in B\setminus A$.  We say a graph $G$ is {\em
$k$-complemented} if $\mu_G(u,v) \ge k) \implies uv \in E(G)$ holds for
every two vertices $u,v\in V$. We may canonically modify the input graphs $G$ and $H$ \lgi, by
adding these additional edges and making them $k$-complemented.

\begin{theorem}[\cite{iso_fpt_treewidth}, Theorem~5.5] \label{thm:pruned_bags}
Let $k$ be a positive integer, and let $G$ be a graph on $n$ vertices that is connected and
$k$-complemented. There exists an algorithm that computes in time $2^{\O(k^5\log k)}\cdot n^3$ an
isomorphism-invariant family of bags $\mathcal{B}$ with the following properties:
\begin{enumerate}
  \item $|B|\le \O(k^4)$ for each $B\in\mathcal{B}$,
  \item $|\mathcal{B}|\le 2^{\O(k^5\log k)}\cdot n^2$,
  \item Assuming $\tw(G)<k$, the family $\mathcal{B}$ captures some tree
        decomposition of $G$ that has width $\O(k^4)$.
  \item The family $\mathcal{B}$ is closed under taking subsets.
\end{enumerate}
\end{theorem}

\begin{theorem} \label{thm:fpt_treewidth}
The problem \lgi can be solved in \cFPT time $2^{\O(k^5 \log k)} n^5$ where $k = \tw(G)$.
\end{theorem}

\begin{proof}
We use the algorithm of Lemma~\ref{lem:xp_treewidth}, where $C_i$'s and $D_i$'s are from the
collection $\calB$ of Theorem~\ref{thm:pruned_bags}.  The total number of pairs $(C_i,C_i^j)$ and
$(D_{i'},D_{i'}^{j'})$ is bounded by $2^{\O(k^5 \log k)} n^3$~\cite[p. 20]{iso_fpt_treewidth}.  The
dynamic programming in~\cite[Theorem 6.2]{iso_fpt_treewidth} is done according to the potential
function $\Phi(D_{i'},D_{i'}^{j'}) = 2|D_{i'}^{j'}| + |D_{i'}|$. We use nice tree decompositions,
so in each step, the dynamic programming either introduces a new node into the bag $D_{i'}$, or
moves a node from the bag $D_{i'}$ to $D_{i'}^{j'}$, or joins several pairs with the same bag
$D_{i'}$. In all these operations, we check existence of a list-compatible isomorphism, using
dynamic programming, exactly as in Lemma~\ref{lem:xp_treewidth}.\qed
\end{proof}

\section{Conclusions} \label{sec:conclusions}

We conclude this paper with description of related results and open problems.

\heading{Forbidden Images.}
We note that Lubiw~\cite{lubiw1981some} used a different definition of \lgi: for every vertex $u \in
V(G)$, we are given a \emph{list of forbidden images} $\calF(u) \subseteq V(H)$ and we want to find
an isomorphism $\pi : G \to H$ such that $\pi(u) \notin \calF(u)$. The advantage of forbidden lists
is that we can express \gi in space $\O(n+m)$, but the input for \lgi is of size $\O(n^2)$. On
the other hand, we consider lists of allowed images more natural (for instance, list coloring is
defined similarly) and also such a definition appears naturally in~\cite{fkkn}. Both statements are
clearly polynomially equivalent, and the main focus of our paper is to distinguish between tractable
and intractable cases for \lgi.

\heading{Group Reformulation.}
Luks~\cite{luks1982isomorphism} described the following group problem which
generalizes computing automorphism groups of graphs. Let $\Omega$ be a ground
set and let $\Gamma$ be a group acting on $\Omega$.  Further, let $\Omega$ be
colored. We want to compute the subgroup of $\Gamma$ which is color preserving.
When $\Gamma$ is the symmetric group acting on all pairs of vertices $V(G)$
which are colored by two colors (corresponding to edges and non-edges in $G$),
then the computed subgroup is $\Aut(G)$. To generalize graph isomorphism in
this language, we have two colorings and we want to find a color-preserving
permutation $g \in \Gamma$. We note that Babai~\cite{babai_quasipoly} calls
these generalization as the \emph{string automorphism/isomorphism problems}.

A similar generalization of \lgi was suggested to us by Ponomarenko. We are given a group
$\Gamma$ acting on a ground set $\Omega$ and for every $x \in \Omega$, we have a list $\L(x)
\subseteq \Omega$. We ask whether there exists a permutation $g \in \Gamma$ such that $g(x) \in
\L(x)$ for every $x \in \Omega$. We obtain \lgi either similarly as above, or when $\Omega = V(G)$
and $\Gamma = \Aut(G)$.

We may interpret our results for \lgi using this group reformulation.  The robust combinatorial
algorithms work because the groups $\Gamma$ are highly restricted. In particular, for trees,
Jordan~\cite{jordan1869assemblages} proved that $\Aut(G)$ is formed by a series of direct products
and wreath products with symmetric groups, to it has a tree structure. Therefore, the algorithm of
Theorem~\ref{thm:trees} solves \lgi on $\Aut(G)$ by a bottom-up dynamic algorithm.  Similar
characterizations were recently proved for interval, permutation and circle graphs~\cite{kz,kz15}
and for planar graphs~\cite{knz}, and are used in the algorithms of Theorems~\ref{thm:planar},
\ref{thm:int}, \ref{thm:perm}, and~\ref{thm:circle}. For graphs of bounded genus or bounded
treewidth, no such detailed description of the automorphism groups is yet known, but they are likely
restricted as well. On the other hand, for cubic graphs, the automorphism groups $\Aut(G)$ may be
arbitrary, so this approach fails, and actually \lgi is \cNP-complete by Theorem~\ref{thm:3reg_npc}.

To attack \lgi from the point of this group reformulation, instead of different graph classes, we
may study it for different combinations of $\Gamma$ and the lists $\L$. First, for which
groups $\Gamma$, it can be solved efficiently for all possible lists $\L$? Second, for which lists
$\L$, the problem can be solved efficiently? We did not try to attack the problem much
in the second direction, aside Lemma~\ref{lem:list_size_two} and Theorem~\ref{thm:3reg_npc}. For
instance, when $\L$ is a partitioning of $\Omega$, the problem is easy since we get the usual
color-preserving isomorphism problem.

\heading{$\boldsymbol k$-dimensional Weisfieler-Leman refinement ($\boldsymbol k$-WL).} The classical
$2$-WL~\cite{wl,weisfeiler} colors vertices of a graph and it initiates with different colors for
each degree. In each step, it takes vertices of one color class and partitions them by different
numbers of neighbors of other color classes. It stops when no partitioning longer occurs. Its
generalization $k$-WL~\cite{babai77,immerman_lander} colors and partitions $(k-1)$-tuples according
to their adjacencies.

Certainly, when $G$ and $H$ are isomorphic graphs, they are partitioned and
colored the same.  So when $G \not\cong H$, $k$-WL distinguishes $G$ and $H$,
for a suitable value of $k$. If we prove for a graph class that $k$ is small
enough, we obtain a combinatorial algorithm for \gi.  For instance,
Grohe~\cite{grohe_graph_structure} proves that for every graph $X$, there
exists a value $k$ such that two $X$-minor free graphs are either isomorphic,
or distinguished by $k$-WL.  This does not translate to \lgi since $k$-WL
applied on $G$ only estimates the orbits of $\Aut(G)$.  When $G \cong H$, and
we may test this, assuming that \gi can be decided efficiently for $G$ and $H$,
we obtain two identical partitions. They may be used to reduce sizes of the
lists, but we still end up with the question whether there exists a
list-compatible isomorphism. In Section~\ref{sec:group_theory}, we show that it
is \cNP-complete to decide \lgi even when sizes of all lists are bounded by 3.

\heading{Bounded Rankwidth.} Rankwidth generalizes treewidth in the way that bounded treewidth
implies bounded rankwidth, but not the opposite. Very recently, the first \cXP algorithm for graph
isomorphism of graphs of bounded rankwidth was described~\cite{bounded_rankwidth}.  The approach is
by computing automorphism-invariant tree decomposition (which should translate to \lgi), but then
further group theory is applied to test whether the decompositions are isomorphic.  It is a very
interesting question whether group theory can be avoided and the problem can be solved in a purely
combinatoric way. Therefore, determining the complexity of \lgi for graphs of bounded rankwidth is
one of major open problems and it might give an insight into this question as well.

Also, rankwidth is closely related to cliquewidth: when one parameter is bounded, the other is
bounded as well. The graphs of cliquewidth at most 2 are called \emph{cographs} and can be
represented by \emph{cotrees}. Their isomorphism can therefore be reduced to isomorphism of cotrees
and solved in polynomial time~\cite{cographs}, and this approach should translate to \lgi. Very
recently, a combinatorial polynomial-time algorithm for graph isomorphism of graphs of cliquewidth
at most 3 was described~\cite{cliquewidth_three} which might translate to \lgi as well.

\heading{Excluded Minors.} Another major open problem for \lgi is its complexity for graphs with
excluded minors. As described in Introduction, the original polynomial-time algorithm of
Ponomarenko~\cite{ponomarenko_minor} for \gi is based heavily on group theory, and his technique
unlikely translates to \lgi. But it seems doubtful that the problem will be \cNP-complete, since
new combinatorial structural and algorithmic results may be applied.

Robertson and Seymour~\cite{robertson_seymour} proved that every graph $G$ with an excluded minor
can be decomposed into pieces which are ``almost embeddable'' to a surface of genus $g$, where $g$
depends on the minor. The recent book of Grohe~\cite{grohe_graph_structure} describes the seminal
idea of automorphism-invariant treelike decompositions. A \emph{treelike decomposition} generalizes
a classical tree decomposition by replacing a tree of bags by a directed acyclic graph of bags.
Unlike tree decompositions, a treelike decomposition $D$ of $G$ can be constructed with two
additional properties. Firstly, it is \emph{automorphism-invariant}, meaning that every automorphism of $G$
induces an automorphism of $D$. Secondly, it is \emph{canonical}, meaning that for two isomorphic graphs
$G$ and $G'$, isomorphic treelike decompositions $D$ and $D'$ are constructed. 

The main structural result of Grohe~\cite{grohe_graph_structure} is that every graph $G$ with an
excluded minor has a canonical automorphism-invariant treelike decomposition for which the graphs
induced by bags called \emph{torsos} are ``almost embeddable'' to a surface of genus $g$, where $g$
depends on the minor. Therefore, to solve \lgi on graphs with excluded minors, we need to prove the
following:
\begin{packed_enum}
\item We need to show that \lgi can be solved on almost embeddable graphs in polynomial time. We may
use the results of Theorems~\ref{thm:bounded_genus} and~\ref{thm:fpt_treewidth} to do so.
\item We need to prove Lifting Lemma for \lgi, stating the following: If we can compute a canonical
automorphism-invariant treelike decomposition of a class $\calC$ in polynomial time and we can solve
\lgi for its torsos $\calA$ in polynomial time, then we can solve \lgi for $\calC$ in polynomial
time as well. Here, we might modify the algorithm for lifting of canonization by
Grohe~\cite{grohe_graph_structure}.
\end{packed_enum}

We note that it is quite difficult to understand and describe everything in detail. The book of
Grohe~\cite{grohe_graph_structure} is very extensive (almost 500 pages) and described in the
language of graph logics. Unfortunately, no purely combinatorial description of the results is
available, and we believe that such a description of a combinatorial algorithm for solving graph
isomorphism of graphs with excluded minors would be desirable.  A combinatorial description of
treelike decompositions is described by Grohe and Marx~\cite{grohe_marx}, but an algorithm for the
graph isomorphism problem of graphs with excluded minors is only used as a black box.

\heading{Bounded Eigenvalue Multiplicity.} The polynomial-time algorithms for \gi of graphs of
bounded eigenvalue multiplicity~\cite{babai_eig_multiplicity,ponomarenko_eig_multiplicity} are
heavily based on group theory. Actually, already the case of multiplicity one is non-trivial. It
seems unlikely that these results will translate to \lgi, but constructing an \cNP-hardness
reduction with bounded eigenvalue multiplicity seems non-trivial.

\heading{Forbidden Subgraphs, Induced Subgraphs and Induced Minors.} There are several papers
dealing with \gi for classes of graphs with excluded subgraphs, induced subgraphs and induced
minors, and again the question is which results translate to \lgi. Otachi and
Schweitzer~\cite{otachi_schweitzer_subgraph} prove a dichotomy for excluded subgraphs. The
\cGI-complete cases translate by Theorem~\ref{thm:vertex_gadget_reductions}, but the polynomial
cases follow from~\cite{grohe_marx} which does not seem to translate. More complicated
characterizations are known for forbidden induced
subgraphs~\cite{booth_colbourn_GI,kratsch_schweitzer,schweitzer_dichotomy}.  Belmonte et
al.~\cite{belmonte_otachi_schweitzer} describe dichotomy for forbidden induced minors.

\heading{Logspace Results.} For some graph classes, \gi is known to be solvable in \cLog and
other subclasses of \cP. It is a natural question to ask whether these results translate to \lgi.
For instance, graph isomorphism of trees~\cite{tree_log_iso} can be solved in \cLog, with a
similar bottom-up procedure as in the proof of Theorem~\ref{thm:trees}. The celebrated result of
Reingold~\cite{reingold}, stating that undirected reachability can be solved in \cLog, allowed
many other graph algorithms to be translated to \cLog. In particular, \gi is known to be solvable
in \cLog for planar graphs~\cite{planar_3conn_log_iso,planar_log_iso},
$k$-trees~\cite{ktree_log_iso}, interval graphs~\cite{int_log_iso}, and bounded treewidth
graphs~\cite{treewidth_log_iso}.

\bibliographystyle{plain}
\bibliography{list_isomorphism}

\end{document}